\documentclass[aip, reprint, pop, groupedaddress]{revtex4-1}
\usepackage{times}		
\usepackage{booktabs}
\usepackage{placeins}

\usepackage{amssymb,amstext,amsmath, epsf}
\usepackage{graphicx,graphics,epsfig,wrapfig}
\usepackage{listings}
\usepackage{float}
\usepackage{esint}

\usepackage{color}
\usepackage[english]{babel}
\usepackage[T1]{fontenc}
\usepackage[utf8]{inputenc}
\usepackage[font=small,labelfont=bf]{caption}
\usepackage{mwe}    
\usepackage{subcaption}
\usepackage[normalem]{ulem}

\def\prl{{Phys. Rev. Lett.}}

\def\pre{{Phys.\ Rev.\ E}}

\def\pop{Phys. Plasmas}
\def\pof{Phys. Fluids}

\def\jfm{J. Fluid Mech.}
\def\pr{Phys. Rev.}

\def\jcp{J. Comput. Phys.}

\usepackage[normalem]{ulem}

\begin{document}

\title{Ion Species Stratification Within Strong Shocks in Two-Ion Plasmas}

\author{Brett D. Keenan}
\email{keenan@lanl.gov}
\author{Andrei N. Simakov}
\author{William T. Taitano}
\author{Luis Chac\'{o}n}
\affiliation{Los Alamos National Laboratory, Los Alamos, NM 87545, USA}


\begin{abstract}
Strong collisional shocks in multi-ion plasmas are featured in many environments, with Inertial Confinement Fusion (ICF) experiments being one prominent example. Recent work  [Keenan {\it et al.}, \pre \ {\bf 96}, 053203 (2017)] answered in detail a number of outstanding questions concerning the kinetic structure of steady-state, planar plasma shocks, e.g., the shock width scaling by Mach number, $M$. However, it did not discuss shock-driven ion-species stratification (e.g., relative concentration modification, and temperature separation). These are important effects, since many recent ICF experiments have evaded explanation by standard, single-fluid, radiation-hydrodynamic (rad-hydro) numerical simulations, and shock-driven fuel stratification likely contributes to this discrepancy. Employing the state-of-the-art Vlasov-Fokker-Planck code, iFP, along with multi-ion hydro simulations and semi-analytics, we quantify the ion stratification by planar shocks with arbitrary Mach number and relative species concentration for two-ion plasmas in terms of ion mass and charge ratios. In particular, for strong shocks, we find that the structure of the ion temperature separation has a nearly universal character across ion mass and charge ratios. Additionally, we find that the shock fronts are enriched with the lighter ion species, and the enrichment scales as $M^4$ for $M \gg 1$.
\end{abstract}

\maketitle

\section{Introduction}
\label{s:intro}

Inertial Confinement Fusion (ICF) aims to generate nuclear fusion energy via compression and heating of a small capsule filled with thermonuclear fuel. Despite the fact that the fuel is largely in a multi-component plasma state during critical stages of the implosion, single-fluid, radiation-hydrodynamic (rad-hydro) simulations are typically employed to design ICF targets and predict experimental outcomes. For years, ICF experiments have failed to produce results congruent with rad-hydro predictions. Experimentally observed and simulated implosion parameters, such as nuclear yield and capsule compression, often disagree. One puzzling result was obtained in experiments testing hydrodynamic equivalence in ICF implosions. Under the assumption that the single-fluid approximation is adequate, fuel mixtures which have the same total mass and particle number densities should be compressed in a similar way, resulting in a corresponding scaling of the yield with the fuel composition. However, experiments conducted on the OMEGA laser facility found an anomalous scaling. For example, a capsule filled with a 50/50 atomic mixture of D-$^3$He has been observed to produce about a half of the yield predicted by scaling from pure D$_2$ and nearly pure $^3$He. \citep{rygg06, rinderknecht15} 

The validity of simulating capsule implosions with single-fluid hydrodynamics depends critically upon the magnitude of the Knudsen number, $N_K$, the ratio of the constituent ions' mean free path to a characteristic gradient length scale (e.g., the capsule radius). Experimental evidence from the NIF and OMEGA facilities \citep{rosenberg14a} demonstrates that the discrepancy between rad-hydro and experimental yield is most prominent for large Knudsen numbers (i.e.\ $N_k \sim 1$) -- exceeding a factor of $100$ for an equimolar D-$^3$He mixture when $N_K \approx 9$. Conversely, high-density, short mean free path setups (i.e.\ those for which $N_k \ll 1$) produce experimental yields which agree much better with rad-hydro simulations. \citep{rosenberg14, rosenberg14a, rosenberg15, rinderknecht15} Large Knudsen numbers result in two important effects. Firstly, with increasing $N_K$, the behavior of individual ion species will deviate from that predicted by a single-fluid description. This is believed to be the result of differential mass diffusion of various ion species \citep{rosenberg14a, rinderknecht14}, and ion temperature separation among the species. \citep{rinderknecht15} Both types of behavior have been experimentally observed in moderately large Knudsen number regimes. Secondly, kinetic effects -- which are not captured by hydrodynamic simulations -- may be important.

Plasma shocks, which are normally used to drive ICF capsule implosions, are prime examples of large $N_K$ systems. Indeed, according to simple hydrodynamic estimates, as the Mach number, $M$, of a collisional shock increases, so should the Knudsen number. \citep{zeldovich67} Thus, the hydrodynamic description of a collisional shock is only formally valid for small Mach numbers ($M - 1 \ll 1$), where $N_K \sim 2(M-1)$. \citep{jukes57, mott-smith51} A kinetic formalism should be employed for moderate and strong shocks. Nonetheless, most of the studies to date have focused on the structure of a planar shock in a single species plasma using hydrodynamic formalism. \citep{jukes57, shafranov57, jaffrin64, grewal73}

Moreover, the detailed structure of strong, collisional shocks in multi-ion plasmas has been largely unexplored. Several Vlasov-Fokker-Planck (VFP) simulations of binary-ion plasma shocks, in planar and spherical geometry, were conducted using the FPION code for an equimolar mixture of deuterium and helium-3. \citep{larroche12} Unfortunately, FPION simulations have failed to explain the experimental discrepancies in ICF implosions for the high Knudsen number regime. \citep{larroche12, larroche16} The electromagnetic Particle-in-Cell (PIC) code, LSP, has also been used to explore ICF implosions, \citep{le16} and kinetic shocks in particular. \citep{bellei14, bellei14a}

By employing another VFP code, iFP, and using comparisons with multi-ion hydro simulations and semi-analytics, we showed \citep{keenan17} that neither FPION \citep{larroche12} nor LSP \citep{bellei14} correctly predicted the detailed kinetic structure of a strong ($M \gg 1$), planar, D-$^3$He plasma shock. iFP is a state-of-the-art code, which is mass, energy, and momentum conserving; it is also adaptive and well verified. \citep{taitano15, taitano16, yin16, taitano17, taitano17a} iFP treats ions fully kinetically, resolving all ion species within their own separate velocity-spaces, while simultaneously solving the quasi-neutral fluid equations for electrons. \citep{simakov14}  Additionally, our multi-ion hydro code and theory are grounded in a multi-species generalization of the Braginskii equations. \citep{simakov14, simakov16, simakov16a, simakov17}

While the kinetic structure of a multi-ion plasma shock remains a topic of ongoing research, a number of basic features are known. For example, the basic structure of any plasma shock is a function of the Mach number. There are three principal regimes: 1) the weak shock limit ($M - 1 \ll 1$), which admits an analytical treatment; \citep{simakov17} 2) an intermediate regime ($M - 1 \sim 1$); and 3) the strong-shock limit ($M \gg 1$). The weak limit is characterized by extremely weak ion species separation in concentration $\propto (M-1)^2$ and temperature $\propto (M-1)^3$. The hydrodynamic description is formally valid here, and the ions and electrons have approximately the same temperature. The intermediate regime is characterized by some appreciable ion stratification, as well as full thermal decoupling of electrons from the ions. In this regime, kinetic effects exist, but they are not prominent enough to cause considerable deviation from hydrodynamic predictions. \citep{keenan17} 

\begin{figure}
\includegraphics[angle = 0, width = 1\columnwidth]{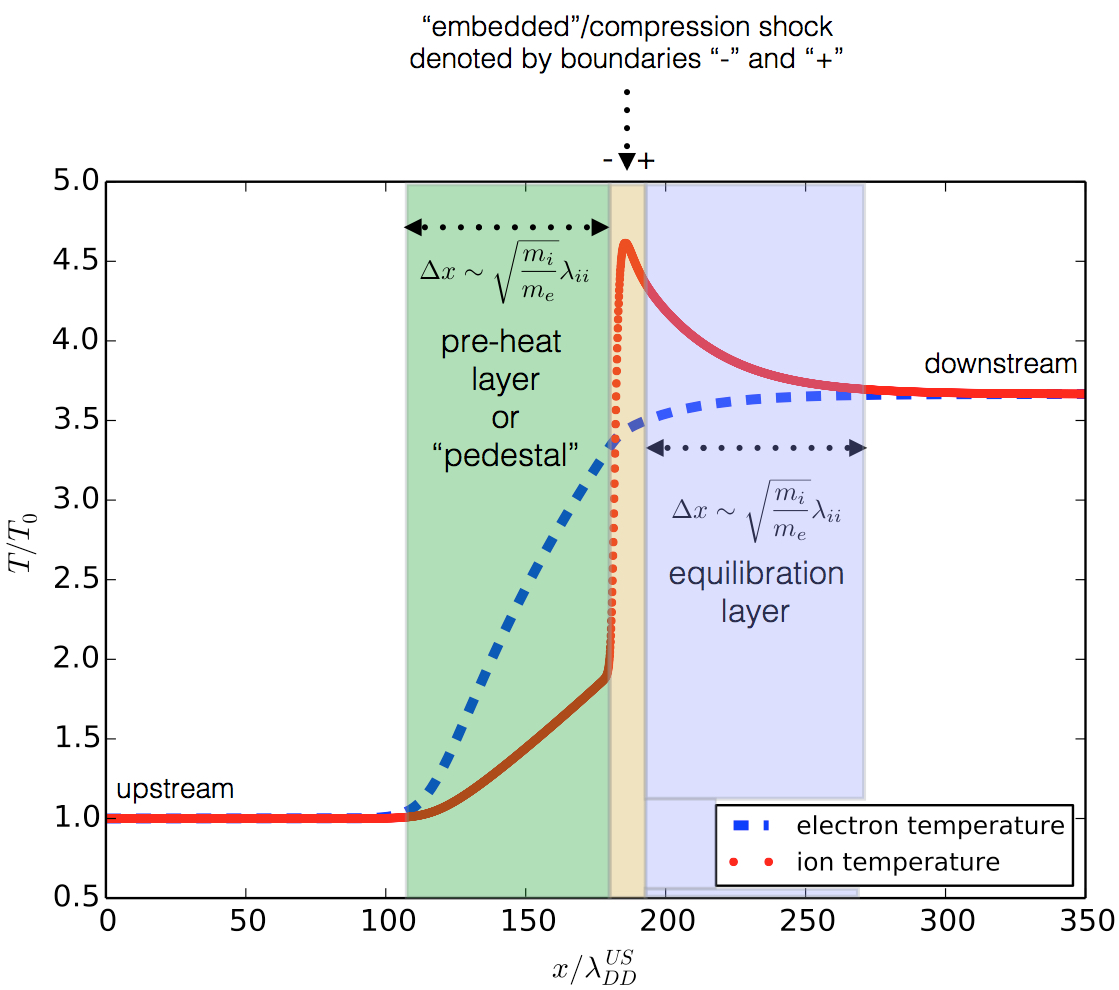}
\vskip-0.1cm
\caption{(Color online). Normalized temperature profiles for a hydro plasma shock with $M \gg 1$. $T_0$ is the upstream temperature. This figure was obtained from our previous work. \citep{keenan17}}
\label{shock_diagram}
\end{figure}

Strong shocks are characterized by order-unity ion concentration and temperature stratification. The structure of a strong, single-ion species, hydro plasma shock, unlike the kinetic equivalent, is well known. \citep{jukes57, shafranov57, jaffrin64, grewal73} As depicted in Fig.\ \ref{shock_diagram}, a strong hydro shock is divided into three principal domains: 1) an electron pre-heat ``pedestal'' region where the electron temperature far exceeds the ion temperature; 2) the embedded (or compression) ion shock where the plasma density increases by a factor of about four (for the adiabatic index, $\gamma = 5/3$); and 3) the equilibration layer where the electrons and ions relax to the same downstream temperature. Both regions 1) and 3) extend over lengths of order $\sqrt{\frac{m_i}{m_e}}\lambda_{ii}$ (where $m_i$ and $m_e$ are the ion and electron masses, respectively, and $\lambda_{ii}$ is the downstream ion-ion mean free path); whereas 2) is a few ion-ion mean free paths (mfps) long. This basic structure is largely the same when shocks are treated kinetically. \citep{keenan17}
\newline
\indent
In general, binary plasma shocks should be fully described by four dimensionless parameters: 1) The Mach number, $M$; 2) the initial concentration (or mass fraction) of species one (which is, by convention, the lightest species), $c_0 \equiv \rho_1^0/\rho_0$, where $\rho_1^0$ and $\rho_0$ are the upstream species one and the total ion mass densities, respectively; 3) the ion mass ratio, $\mu \equiv m_2/m_1$, where $m_1$ and $m_2$ are the species one and two ion masses, respectively; and 4) the ion charge ratio, $\xi \equiv Z_2/Z_1$, where $Z_1$ and $Z_2$ are the ion species one and two charges, respectively. 

Previously, \citep{keenan17} we only considered D-$^3$He ($\mu = 3/2$; $\xi = 2$) plasma shocks across ($M$, $c_0$). In this work, we continue our study of steady-state, multi-ion, plasma shocks with iFP by investigating the complete parameter space, $p \in (M, c_0, \mu, \xi)$. In particular, we quantify ion species separation in temperature and concentration as functions of the Mach number and initial concentration for various plasma mixtures (varying $\mu$ and $\xi$). For the first time, we quantify the kinetic enhancement of mass diffusion and temperature separation in the large Mach number regime. For simplicity, we once again consider a binary-ion plasma mixture in planar geometry.

The paper is organized as follows. In Section \ref{s:enrich}, we study how the shock enriches the upstream region with a higher concentration of the lightest species. We consider this effect vs.\ the Mach number, initial concentration, and for various plasma mixtures. Next, in Section \ref{s:temp_sep}, we uncover the physics that underlies temperature separation between ion species. Once more, we consider this effect vs.\ the Mach number, initial concentration, and for various plasma mixtures. Finally, we conclude in Section \ref{s:disc}. 
 
 \FloatBarrier
 
\section{Differential Mass Diffusion in Multi-Ion Plasma Shocks}
\label{s:enrich} 

Particles respond to gradients in hydro quantities (temperature, density, etc.) in ways that are differentiated by mass and charge. In particular, differential diffusion of the lighter ion species in a binary-ion plasma mixture can lead to an enhancement of the lighter species concentration, $c \equiv \rho_l/\rho$ (where $\rho_l$ is the mass density of the lighter species, and $\rho$ is the total ion mass density). With $N_K \ll 1$, the change in the lighter species concentration (from its upstream value, $c_0$) is given by: \citep{kagan12, simakov17}  
\begin{equation}
\begin{split}
 c - c_0 = \hat{D}_3\left[\frac{\text{d}c}{\text{d}\hat{x}} + \kappa_P\frac{\text{dlog} \ \hat{p}_i}{\text{d}\hat{x}} + \kappa_{Ti}^{(i)}\frac{\text{dlog} \ \hat{T}_i}{\text{d}\hat{x}} \right. \\
\left. +  \kappa_E\frac{\hat{T}_e}{\hat{T}_i}\frac{\text{dlog} \ \hat{p}_e}{\text{d}\hat{x}}\ + \left(\kappa_T^{(e)} + \kappa_E\frac{\hat{T}_e}{\hat{T}_i}\beta_0\right)\frac{\text{dlog} \ \hat{T}_e}{\text{d}\hat{x}}\right],
\end{split}
\label{c-c0_def}
\end{equation}
where {\it p} and {\it T} refer to electron, ``$e$'' and ion, ``$i$'' pressures and temperatures (respectively), the $\kappa$ terms are mixture-dependent coefficients, $\hat{D}_3$ is the overall diffusion coefficient, and the ``hats'' denote quantities normalized to their respective upstream values (for details, see Simakov, {\it et al.} \citep{simakov17}). Distance is normalized to the upstream DD mfp, $\lambda_{\text{DD}}^{US}$ (i.e., $\hat{x} \equiv x/\lambda_{\text{DD}}^{US}$). We define the mean free path for collisions of a particle $\alpha$ with a particle $\beta$ as:
\begin{equation}     
\lambda_{\alpha\beta} \equiv \left(\frac{3}{4\sqrt{\pi}}\right)\left(\frac{T^2m_\beta}{Z_\alpha^2 Z_\beta^2 e^4 n_\beta m_r \text{ln}\Lambda_{\alpha \beta}}\right),
\label{lamb_def}
\end{equation}
where
\[
    m_r \equiv 
\begin{cases}
    \frac{m_\alpha m_\beta}{m_\alpha + m_\beta},& \text{if } m_\alpha \neq m_\beta \\
    m_\alpha,              & \text{otherwise}
\end{cases}
\]
$T$ is the temperature (assumed equal for all species), $n_\beta$ is the $\beta$ number density, $eZ_\alpha$ and $eZ_\beta$ are the $\alpha$ and the $\beta$ charges, respectively, and $\text{ln}\Lambda_{\alpha \beta}$ is the Coulomb logarithm. \citep{rinderknecht17} The total mfp for species $\alpha$ is then:
\begin{equation}     
\lambda_\alpha \equiv \left[\sum_\beta \lambda_{\alpha \beta}^{-1}\right]^{-1},
\label{lamb_tot_def}
\end{equation}
where the sum is over all ion species and the electrons. All simulations presented herein assume a constant Coulomb logarithm of 10 for all species. Such an assumption does not produce qualitative changes in the results. 

\FloatBarrier

\subsection{Species Stratification in Shock Fronts}
\label{s:strat_x}

\begin{figure}[thb]
\includegraphics[angle = 0, width = 1\columnwidth]{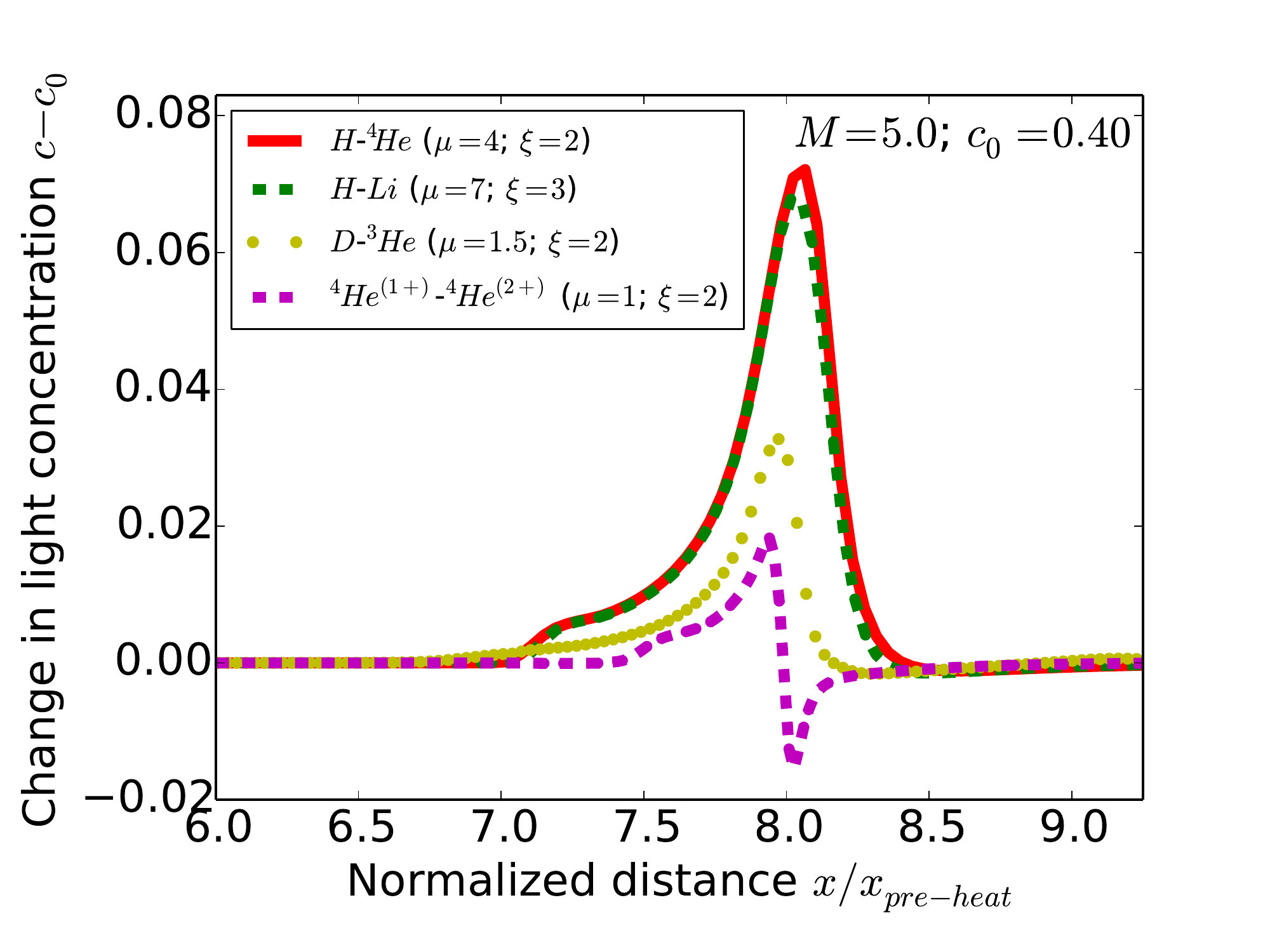}
\vskip-0.1cm
\caption{(Color online). Change in the light ion species concentration for various plasma mixtures with large ion mass ratios, $\mu$ and of varying charge ratio, $\xi$. Distances are normalized to the width of the electron pre-heat layer, $x_{pre-heat}$, for each mixture.}
\label{c-c0_mass}
\end{figure}

\begin{figure}[thb]
\includegraphics[angle = 0, width = 1\columnwidth]{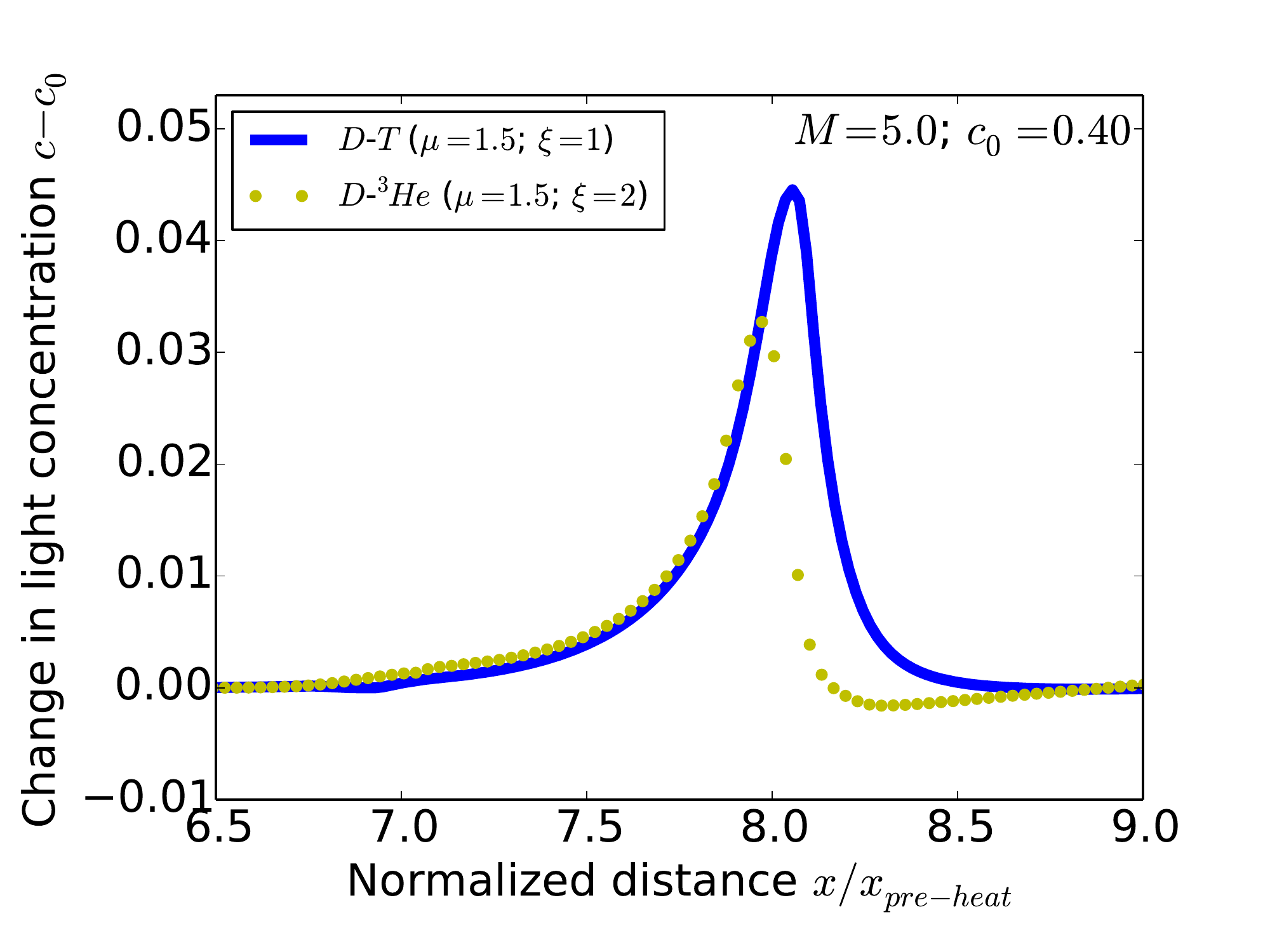}
\vskip-0.1cm
\caption{(Color online). Change in the light ion species concentration for plasma mixtures with different charge ratios, $\xi$.}
\label{c-c0_mZ}
\end{figure}

Figures \ref{c-c0_mass} and \ref{c-c0_mZ} show the spatial variation in the light ion species concentration for various plasma mixtures ($M = 5.0$; $c_0 = 0.40$) from iFP. We normalize distances to the width of the electron pre-heat layer, $x_{pre-heat} \sim \lambda_{ee}\frac{v_{the}}{u_0}$, where $v_{the} = \sqrt{2T_{DS}/m_e}$ is the electron thermal velocity, $\lambda_{ee}$ is the electron-electron mfp, $u_0$ is the shock velocity, and $T_{DS}$ is the downstream temperature (which is the same for all species). Figure \ \ref{c-c0_mass} considers the impact of the ion mass ratio, $\mu$. It is evident, given the similarity between H-Li ($\mu = 7$; $\xi = 3$) and H-$^4$He ($\mu = 4$; $\xi = 2$) in Fig.\ \ref{c-c0_mass}, that the profile is largely determined by the lighter species when the heavier species has a considerably greater mass. Moreover, the peak change in the light species concentration roughly scales with $\mu$, with $^4$He$^{(1+)}$-$^4$He$^{(2+)}$ ($\mu = 1$; $\xi =2$) showing the smallest peak. Additionally, the $^4$He$^{(1+)}$ depletion is principally centered around the equilibration layer of the shock. This suggests that, since the masses of both species are the same, the difference in the self-equilibration time-scales is the most significant contributor to their separation.
\newline
\indent
Holding $\mu$ constant, we explore the impact of just $Z$ in Figure \ref{c-c0_mZ}, which shows D-$^3$He ($\mu = 3/2$; $\xi = 2$) and D-T ($\mu = 3/2$; $\xi = 1$) mixtures. The change in the deuterium concentration for the D-T plasma has a higher peak than for D-$^3$He plasma in Fig.\ \ref{c-c0_mZ}. Given the $Z^2_{^3\text{He}}$-dependence in the D-$^3$He mfp, deuterium is more collisional in a D-$^3$He plasma than it is in a hydro-equivalent D-T plasma (where $Z_{\text{T}}  = Z_\text{D} = 1$), and this is the likely cause for the taller peak seen in Fig.\ \ref{c-c0_mZ}. 
\FloatBarrier

\subsection{Species Stratification vs.\ the Mach Number}
\label{s:strat_M}

\begin{figure}[thb]
\includegraphics[angle = 0, width = 1\columnwidth]{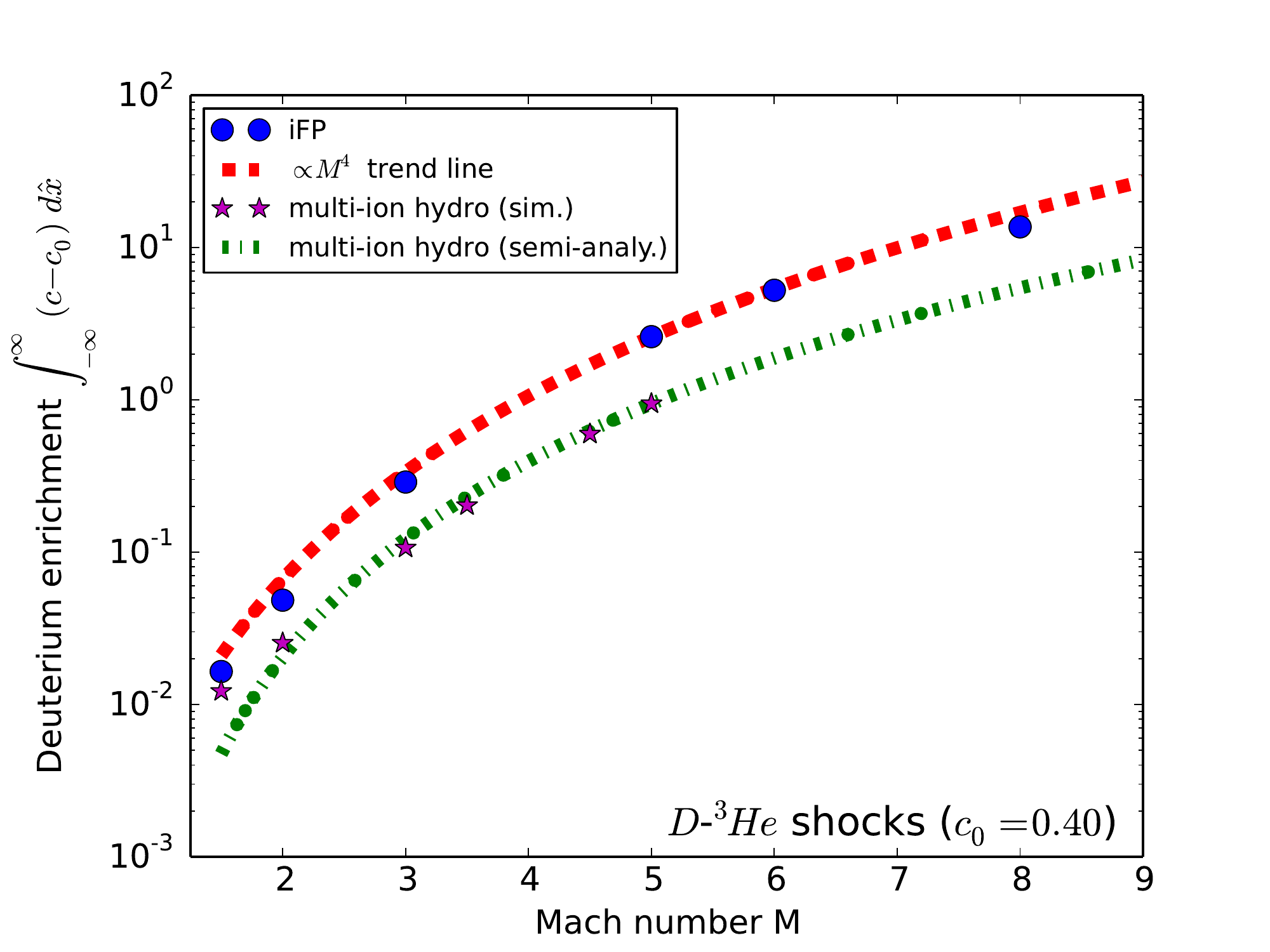}
\vskip-0.1cm
\caption{(Color online). Deuterium enrichment vs.\ Mach number for D-$^3$He shocks with $c_0 = 0.40$.}
\label{c_int_mach_D3He}
\end{figure}

\begin{figure}[thb]
\includegraphics[angle = 0, width = 1\columnwidth]{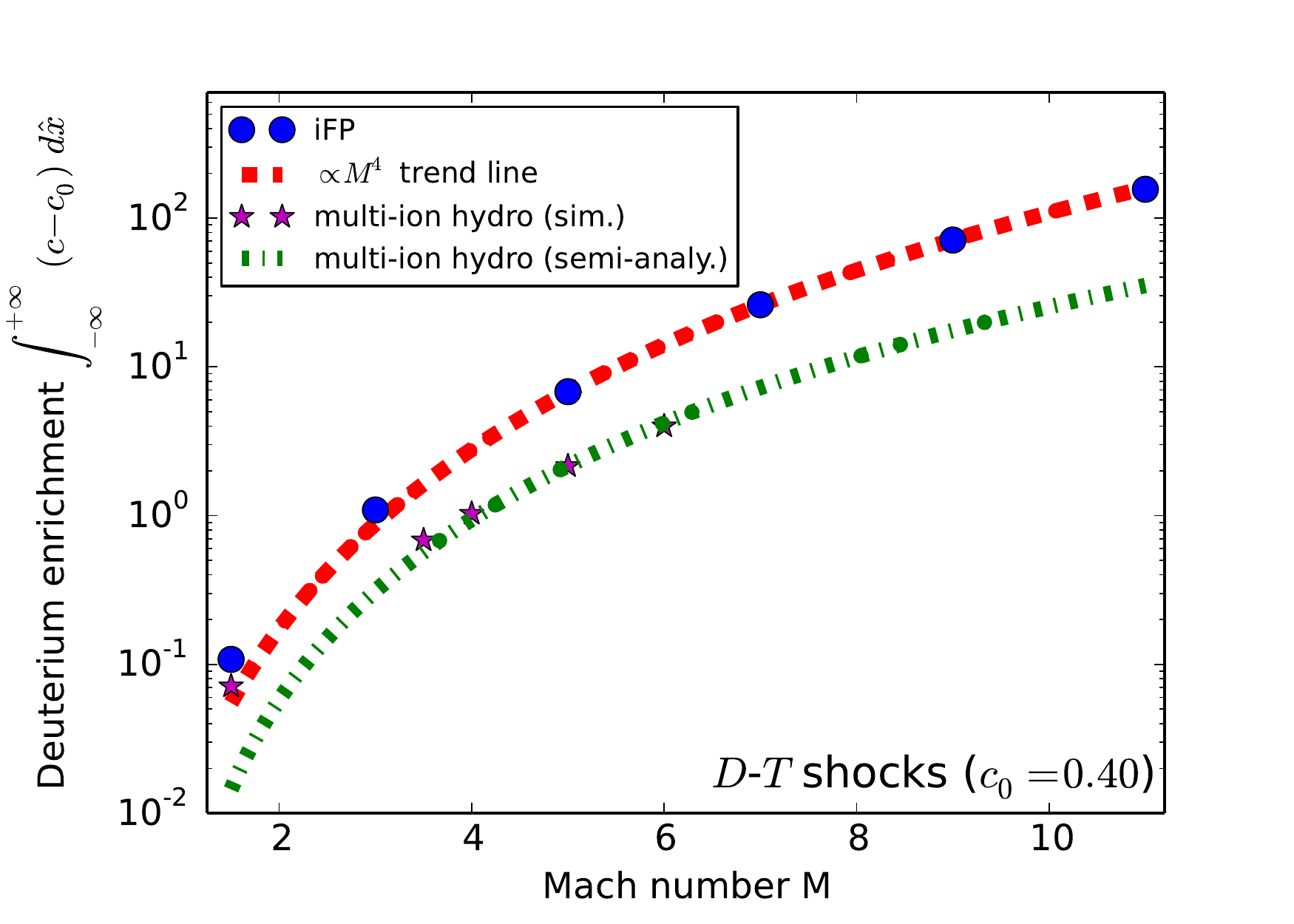}
\vskip-0.1cm
\caption{(Color online). Deuterium enrichment vs.\ Mach number for D-T shocks with $c_0 = 0.40$.}
\label{c_int_mach_DT}
\end{figure}

\begin{figure}[thb]
\includegraphics[angle = 0, width = 1\columnwidth]{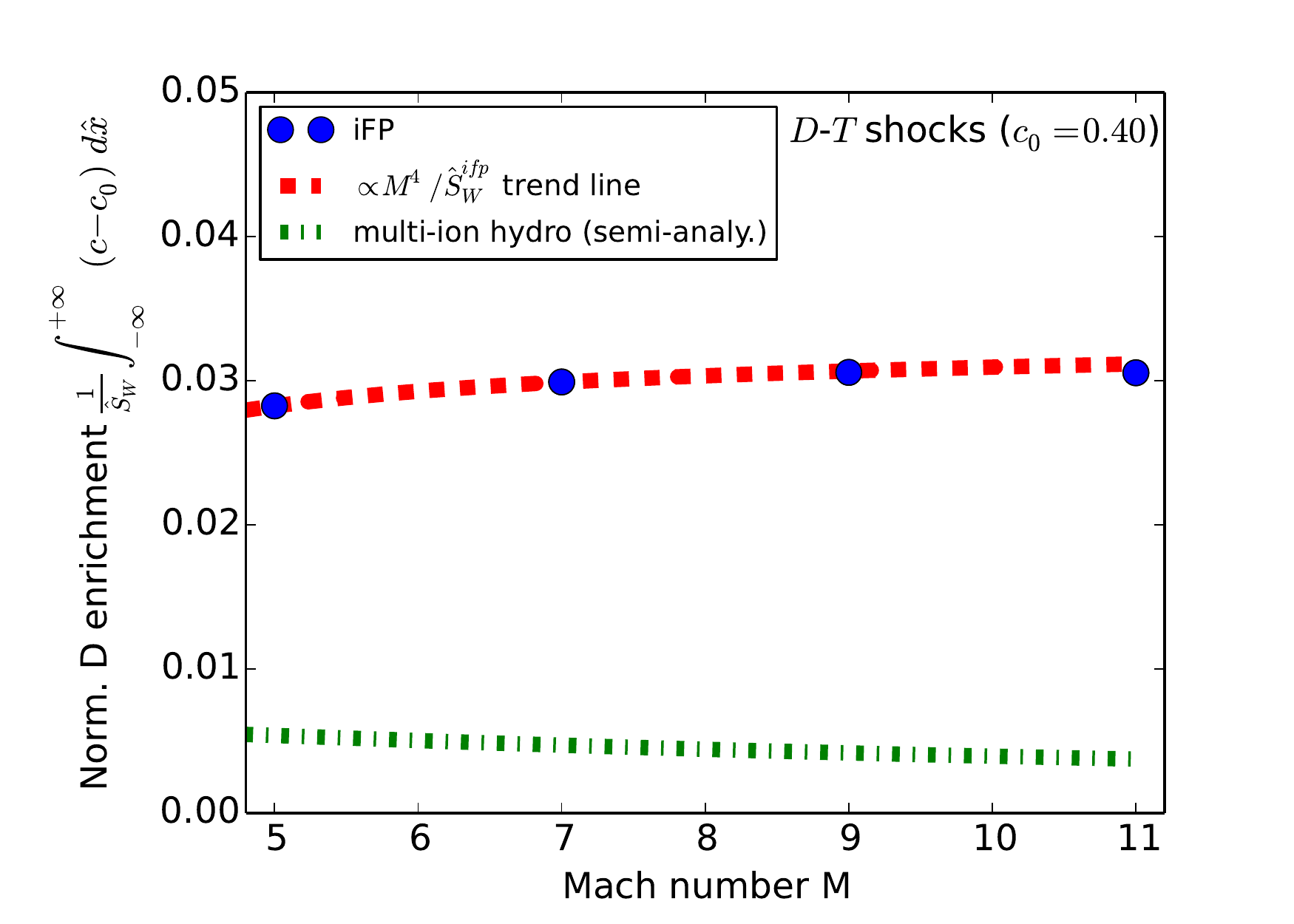}
\vskip-0.1cm
\caption{(Color online). Deuterium enrichment (normalized to the ion shock width, $\hat{S}_W$) vs.\ Mach number for D-T shocks with $c_0 = 0.40$.}
\label{norn_c_int_mach_DT}
\end{figure}

The bulk enrichment of the lighter species is obtained by integrating $c - c_0$ across the entire $\hat{x}$ domain. In Fig.\ \ref{c_int_mach_D3He}, we show the kinetic deuterium enrichment from iFP across a D-$^3$He shock (blue dots) vs.\ Mach number. It is useful to compare the observed trend to the equivalent multi-ion hydro prediction. To this end, we approximately compute the integral of Eq.\ (\ref{c-c0_def}). We note that the $\kappa$ coefficients depend only upon $c_0$, $\mu$, and $\xi$. Furthermore, we take $\hat{T}_e/\hat{T}_i \approx 1$, since the electron and ion temperatures differ only by a factor of a few within the ion compression shock (where most the ion enrichment occurs in a hydro shock). Next, the diffusion coefficient, $\hat{D}_3$, has a Mach number dependence through $\hat{D}_3 \propto \frac{1}{M}\hat{T}_i^{5/2}$. Noting that $\hat{T}_i \propto M^2$ for $M \gg 1$ near the compression shock, and therefore we take $\hat{D}_3 \propto M^4$. The exact coefficient of $\hat{D}_3$ is a function of $c_0$ and the ion mass/charge ratios, but we may eliminate this dependency by shifting the overall curve to a known multi-ion hydro simulation result. With these considerations, we approximate the integral of Eq.\ (\ref{c-c0_def}) as:
\begin{equation}
\begin{split}
\int_{-\infty}^{+\infty} \left(c - c_0\right) \ \text{d}\hat{x} \approx fM^4\left[\kappa_P \ \text{log}\left(\hat{p}_i^1\right) + \kappa_{Ti}^{(i)} \ \text{log}\left(\hat{T}_i^1\right) \right. \\ 
\left. + \kappa_E \ \text{log}\left(\hat{p}_e^1\right) + \left(\kappa_T^{(e)} + \kappa_E\beta_0\right)\text{log}\left(\hat{T}_e^1\right)\right] \\
\propto M^4\text{log}M \approx M^4,
\end{split}
\label{c-c0_int}
\end{equation}
where $f = f(c_0, \mu, \xi)$ is derived from full multi-ion hydro simulations, and $1$ denotes the downstream quantities.
\newline
\indent
In Fig.\ \ref{c_int_mach_D3He}, we plot the multi-ion hydro semi-analytic prediction for the deuterium enrichment (green dash-short-dashed line) from Eq.\ (\ref{c-c0_int}) vs.\ Mach number. Also included in Fig.\ \ref{c_int_mach_D3He} are results obtained from full multi-ion hydro simulations (magenta stars). Our multi-ion hydro simulations are based on the multi-ion generalization of Braginskii's closure scheme. \citep{simakov14, simakov16, simakov16a, simakov17} This scheme includes full multi-ion physics for a two species plasma, except for assuming that both ion species have the same temperature. Temperature separation is a high-order effect in the Knudsen number,\citep{simakov17} and it is essentially kinetic in nature. Figure \ref{c_int_mach_D3He} confirms the kinetic enhancement of D enrichment vs.\ the hydro prediction for $M \gtrsim 2$. 
\newline
\indent 
Figure \ref{c_int_mach_D3He} also confirms that the strong-shock semi-analytic hydro prediction fails for sufficiently low Mach numbers ($M \lesssim 2$), as expected. At $M = 1.5$, the kinetic and multi-ion predictions agree well, as  expected for intermediate-strength shocks. \citep{keenan17} Also note that the kinetic trend closely matches an $M^4$ scaling (red dashed line). Hence, the kinetic enrichment appears to have the same overall dependence upon $M$ as the multi-ion hydro equivalent.
\newline
\indent
We may apply the same analysis to D-T plasmas, as depicted in Fig.\ \ref{c_int_mach_DT}. In this case, effects associated with the charge ratio are eliminated. The results for D-T are qualitatively similar to the D-$^3$He case, but the separation between the kinetic and hydro predictions (for large $M$) appears to be more pronounced. This is consistent with our explanation for the differences in Fig.\ \ref{c-c0_mZ} (i.e., kinetic effects are stronger in the D-T plasma, since its D ions are less collisional than the deuterons in a hydro-equivalent D-$^3$He plasma).
\newline
\indent
Figures \ref{c_int_mach_D3He} and \ref{c_int_mach_DT} show that the deuterium enrichment is significantly enhanced kinetically for strong shocks. This enhancement is more apparent when the deuterium enrichment is normalized to the normalized ion shock width, $\hat{S}_W \equiv S_W/\lambda_{DD}^{US}$. We define the shock width, $S_W$, as the distance over which the total mass density increases from 1.2 times its upstream value, $\rho_0$, to 0.9 times its downstream value, $\rho_1$. This roughly corresponds to the length of the ion compression shock. \citep{grewal73, keenan17} Since $\hat{S}_W$ scales as $M^4$ for strong shocks, \citep{keenan17} the normalized multi-ion hydro deuterium enrichment, $\frac{1}{\hat{S}_W}\int_{-\infty}^{+\infty} \left(c - c_0\right) \ \text{d}\hat{x}$, should approach a finite asymptotic limit as $M \rightarrow \infty$. In Fig. \ref{norn_c_int_mach_DT}, we see that both the kinetic and multi-ion hydro predictions, when normalized to their respective shock widths, have finite asymptotic limits for $M \gg 1$, but the normalized deuterium enrichment for the infinite Mach number limit is an order of magnitude larger than the semi-analytic prediction (which is $\sim 5\ \times \ 10^{-3}$ for $M \rightarrow \infty$).

\FloatBarrier

\subsection{Species Stratification vs.\ Initial Mass Fraction}
\label{s:strat_c0}

The ion enrichment should depend upon, not only the Mach number, but the upstream concentration of the lighter ion species, $c_0$. Ion enrichment as a function of $c_0$ requires a careful treatment, because while $c - c_0$ will necessarily go to zero for $c_0 = 0$ and $c_0 = 1$, the relative change, $(c - c_0)/c_0$, may not. The definition of the relative enrichment must also account for the fact that $\hat{x}  \equiv x/\lambda_{\text{DD}}^{US} \rightarrow 0$ as $c_0 \rightarrow 0$ (since $\lambda_{\text{DD}}^{US} \rightarrow \infty$ as $c_0 \rightarrow 0$). For this reason, we define the relative D enrichment as: $1/(c_0\hat{S}_W)\int (c-c_0)\ \text{d}\hat{x}$; i.e., $(c - c_0)/c_0$ integrated over $\hat{x}$, and then divided by the normalized ion shock width. 

We are offered a clue to the behavior of the relative enrichment near $c_0 = 0$ and $c_0 = 1$ by considering $(c - c_0)/c_0$ at these endpoints. It is apparent that $(c - c_0)/c_0 \rightarrow 0$ for $c_0 \rightarrow 1$. However, $c/c_0 \rightarrow 0/0$ for $c_0 \rightarrow 0$, which is indeterminate. Consequently, the relative enrichment of the lighter species near $c_0 = 0$ may not be zero. 

We provide a theoretical estimate for the relative D enrichment vs.\ $c_0$ as follows. The quantity $c - c_0 = -i/(\rho_0u_0)$, where $i$ is the deuterium mass diffusion flux, \citep{kagan12, simakov17} and the minus sign indicates the direction opposite to the fluid flow (in the shock frame). As depicted in Fig.\ \ref{shock_diagram}, the (hydro) ion temperature near the transition from the ion compression shock to the pre-heat layer has a very steep gradient. This steep gradient is the site of strong kinetic effects, where hot ions (with temperatures on the order of the downstream temperature) counter-stream into the pedestal plasma. \citep{keenan17} Hence, it is reasonable to assume that $i \sim -f(c_0)m_1n_1^1v_{th1}^1$ in the vicinity of the pre-heat region (where $m_1$, $n_1^1$, and $v_{th1}^1$ are the deuterium mass, downstream number density, and downstream thermal velocity, respectively). The unspecified function, $f(c_0)$, was introduced to control any additional $c_0$-dependence on the deuterium mass diffusion flux, and obeys $f(c_0) \rightarrow 0$ for $c_0 \rightarrow 1$, since the relative D enrichment must be zero for $c_0 = 1$.

Multi-ion hydro simulations predict that $c - c_0$ will be very narrowly peaked around the site of the ion compression shock. \citep{keenan17} In contrast, as evidenced by Figs. \ref{c-c0_mass} and \ref{c-c0_mZ}, the kinetic version of $c - c_0$ shows considerable light species enhancement far into the pre-heat region. The majority of the D enrichment occurs a distance, $\Delta{x}$ (computed below) into the pre-heat layer. Thus, given this fact and our estimate for the mass diffusion flux, $\int (c - c_0) dx \sim  f(c_0)m_1n_1^1v_{th1}^1\Delta{x}$. 
\begin{figure}[thb]
\includegraphics[angle = 0, width = 1\columnwidth]{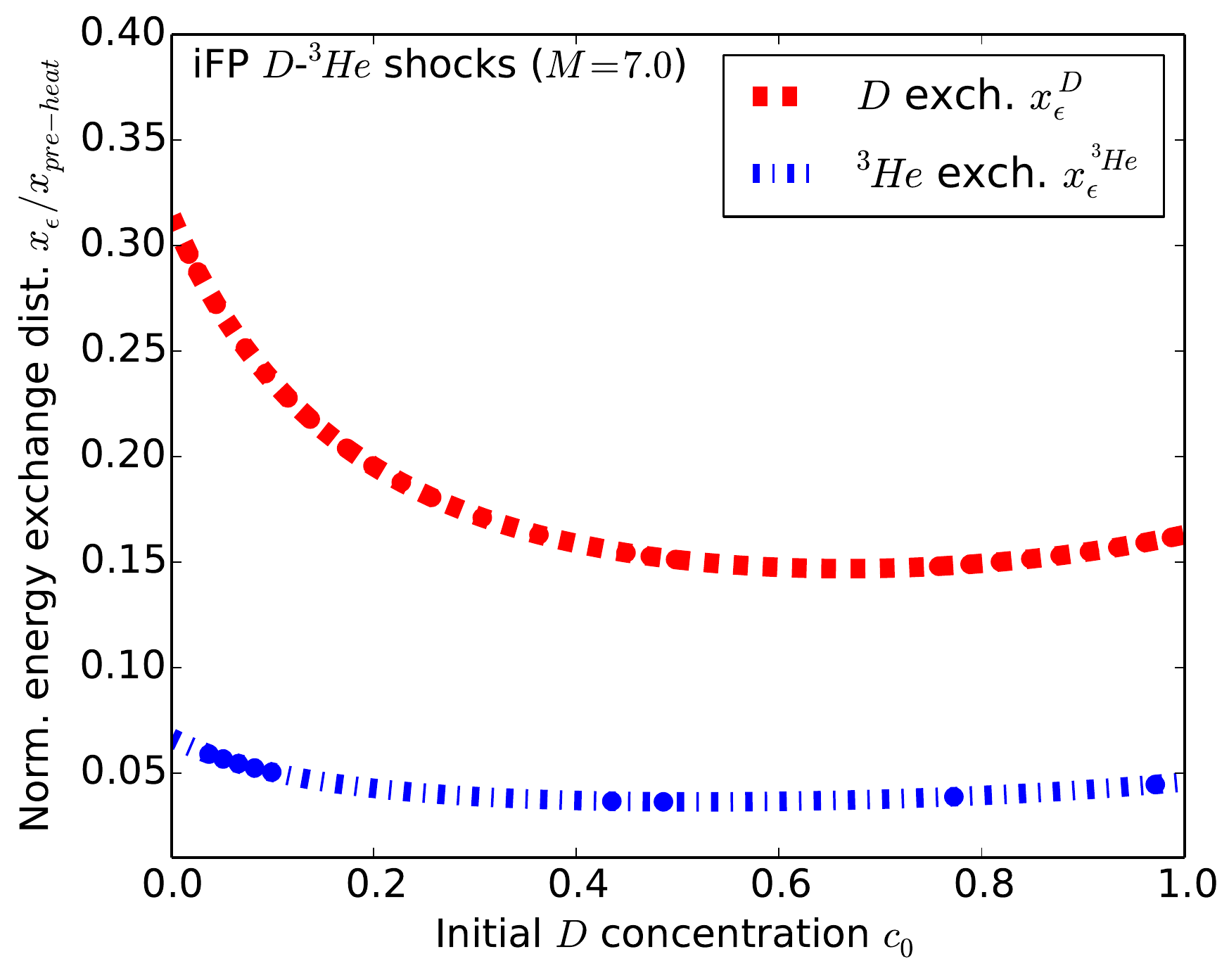}
\vskip-0.1cm
\caption{(Color online). Energy exchange distances vs.\ $c_0$ for D and $^3$He ions within a D-$^3$He shock ($M = 7$). For details on the calculation of the energy exchange distance, see Ref.\ [18].}
\label{erg_exch_comp}
\end{figure}
\newline
\indent
Next, we provide estimates for $f(c_0)$ and $\Delta{x}$. To this end, it is helpful to recall the mechanism whereby a kinetic shock is extended over the hydro one. As discussed in our previous work, \citep{keenan17} the D-$^3$He kinetic shock width is equal to the hydro width plus a correction equal to a characteristic energy exchange distance for kinetic deuterons, $x_\epsilon^D$. It is then reasonable to assume that $\Delta{x} \sim x_\epsilon^\text{D}$. Figure \ref{erg_exch_comp} depicts the energy exchange distances vs.\ $c_0$ for both D and $^3$He ions. The two exchange distances overlap, with $x_\epsilon^{^3\text{He}}$ being the smaller of the two. Within $x_\epsilon^{^3\text{He}}$, the deuterium will be enriched, but the effect will be smaller since kinetic $^3\text{He}$ ions will also be counter-streaming in this region. Hence, the deuterium diffusion flux will have some dependence upon the ratio, $\epsilon_\text{rat} \equiv x_\epsilon^{^3\text{He}}/x_\epsilon^{\text{D}}$; consequently, $f(c_0)$ must be a function of $\epsilon_\text{rat}$. Supposing then that $f(c_0)$ has the simplest form: $f(c_0) = (1 - A\epsilon_\text{rat})$, with $A \equiv x_\epsilon^{\text{D}}(c_0 = 1)/x_\epsilon^{^3\text{He}}(c_0 = 1)$, we have:
\begin{equation}     
\frac{1}{\hat{S}_W}\int_{-\infty}^{+\infty} \left(\frac{c}{c_0}-1\right)\ \text{d}\hat{x} \approx \frac{v_{thD}^1}{\hat{V}_1u_0S_W}\left[x_\epsilon^{\text{D}} - Ax_\epsilon^{^3\text{He}}\right],
\label{rel_enrich_theory}
\end{equation}
where $v_{thD}^1 = \sqrt{2T_{DS}/m_\text{D}}$, and we have used $n_1^1 = c_0\rho_1/m_1$ and $\hat{V}_1 \equiv \rho_0/\rho_1$, where $\rho_1$ is the total downstream mass density. This choice of the constant, $A$, ensures that the relative D enrichment goes to zero as $c_0 \rightarrow 1$. 

\begin{figure}[thb]
\includegraphics[angle = 0, width = 1\columnwidth]{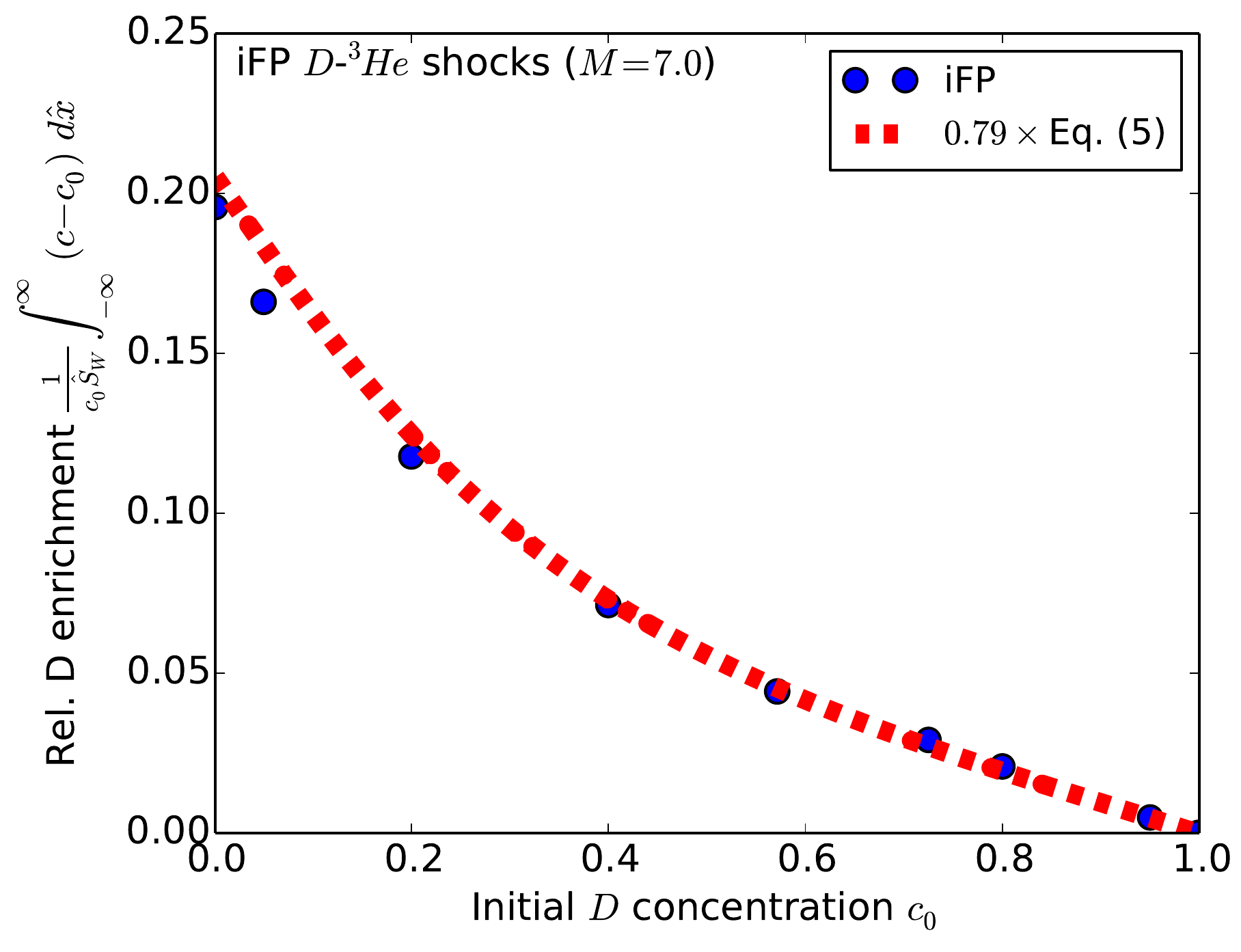}
\vskip-0.1cm
\caption{(Color online). Relative deuterium enrichment vs.\ $c_0$ for D-$^3$He shocks with $M = 7$.}
\label{c_int_c0_D3He}
\end{figure}

Figure \ref{c_int_c0_D3He} displays the relative D enrichment vs.\ $c_0$ from iFP (blue dots). The relative enrichment is maximized for $c_0 \approx 0$, and decreases monotonically to zero at $c_0 = 1$. The analytical theory of weak shocks \citep{simakov17} predicts a similar qualitative behavior. Figure \ref{c_int_c0_D3He} also shows the prediction from Eq.\ (\ref{rel_enrich_theory}) (red dashed line). Remarkably, this simple model captures the majority of the features, including the change in the slope seen near $c_0 = 0.40$ (which is likely due to the slope change in $x_\epsilon^\text{D}$ at that point, as seen in Fig.\ \ref{erg_exch_comp}), as well as the overall amplitude up to a constant multiplicative factor of $0.79$.
\newline
\indent
As an aside, it is worth noting that strong-shock-front enrichment with lighter ion species can have a dramatic dynamical impact at interfaces and singular points (e.g., r = 0, within an ICF capsule). \citep{taitano17b} 

\FloatBarrier

\section{Temperature Separation in Multi-Ion Plasma Shocks}
\label{s:temp_sep} 

Species stratification in shocks also entails temperature separation between ion species. In this Section, we explore in detail the nature of ion temperature separation across the whole shock parameter space, $p \in (M, c_0, \mu, \xi)$. We will first approach this subject from a theoretical perspective.

\subsection{The Theory of Ion Temperature Separation}
\label{s:theory}

The celebrated book on hydro shocks by Zel'dovich and Raizer \citep{zeldovich67} makes the claim that the ion-species temperature within a plasma shock is linearly proportional to the ion mass. Although this claim is made without any justification, it is superficially implied by the Rankine-Hugoniot conditions. However, it is incorrect. To see this, consider the relationship between the total upstream plasma pressure, $P^0$, and the downstream pressure, $P^1$: 
\begin{equation}     
\frac{P^1}{P^0} = \frac{n_e^1T_e^1 + \sum_sn_s^1T_s^1}{n_e^0T_e^0 + \sum_sn_s^0T_s^0} = \left[\frac{2\gamma M^2 - \left(\gamma-1\right)}{\gamma+1}\right].
\label{pressure_jump}
\end{equation}
In the strong-shock limit, $M \gg 1$, we may write:
\begin{equation}     
\frac{n_e^1T_e^1 + \sum_sn_s^1T_s^1}{n_e^0T_e^0 + \sum_sn_s^0T_s^0} \approx 2\left(\frac{u_0^2}{\gamma+1}\right)\frac{n_e^0m_e + \sum_sn_s^0m_s}{n_e^0T_e^0 + \sum_sn_s^0T_s^0},
\label{pressure_jump_limit}
\end{equation}
since $M = u_0/c_s$, where $c_s \equiv \sqrt{\gamma P^0/\rho_0}$ is the upstream sound velocity. Simplifying Eq.\ (\ref{pressure_jump_limit}), we have:
\begin{equation}     
n_e^1T_e^1 + \sum_sn_s^1T_s^1 \approx  2\left(\frac{u_0^2}{\gamma+1}\right)\left(n_e^0m_e + \sum_sn_s^0m_s\right).
\label{pressure_jump_limit1}
\end{equation}
\newline
\indent
Assuming quasi-neutrality, $n_e = \sum_s Z_s n_s$, and noting that: $\rho_0 = m_en_e^0 + \sum_s n_s^0m_s \approx  \sum_s n_s^0m_s$, Eq.\  (\ref{pressure_jump_limit1}) simplifies to:
\begin{equation}     
\left(\sum_s Z_sn_s^1\right)T_e^1 + \sum_sn_s^1T_s^1 \approx  2\left(\frac{u_0^2}{\gamma+1}\right)\sum_sn_s^0m_s,
\label{pressure_jump_limit2}
\end{equation}
which may be expressed as:
\begin{equation}     
\sum_s\left[Z_sT_e^1 + T^1_s - \frac{3}{16}u_0^2m_s\right]n_s^0 \approx  0,
\label{pressure_jump_limit3}
\end{equation}
where we have used the fact that $n_s^1 \approx 4n_s^0$ for $M \gg 1$ (with $\gamma = 5/3$). Since $n_s^0$ is an arbitrary upstream density we have:
\begin{equation}     
Z_sT_e^1 + T^1_s \approx  \frac{3}{16}m_su_0^2,
\label{pressure_jump_limit3}
\end{equation}
which implies that:
\begin{equation}     
T^1_s \approx  \frac{3}{16}m_p\frac{A_su_0^2}{\left[1 + Z_s(T_e^1/T^1_s)\right]},
\label{mass_temp_full}
\end{equation}
where $A_s = m_s/m_p$, and $m_p$ is the proton mass. Finally, for planar shocks, $T_e^1/T^1_s$ is of order unity past the pre-heat layer, \citep{jukes57, shafranov57, jaffrin64, grewal73, keenan17} and thus:
\begin{equation}     
T^1_s \sim  \frac{3}{16}m_p\frac{A_s}{\left(1 + Z_s\right)}u_0^2.
\label{mass_temp}
\end{equation}
Although Eq.\ (\ref{mass_temp}) appears to be generally proportional to $m_pA_s = m_s$, the factor $A_s/(1 + Z_s)$ is typically $\sim 2$ for $Z_s \gg 1$. Consequently, the postshock temperatures for heavy ions, according to Eq.\ (\ref{mass_temp}), will be roughly independent of the ion mass.
\newline 
\indent
 Eq.\ (\ref{mass_temp}) is clearly an oversimplification, since it follows from the Rankine-Hugoniot (jump) conditions, which derive from total mass, energy, and momentum conservation. The jump conditions ignore gradients in fluid quantities (e.g., temperature and density) and plasma transport. However, we may estimate the ion temperature separation directly from fluid equations, provided that we use the correct kinetic closures for heat flux, viscosities, etc. The temperature separation between ion species, $2$ and $1$ (in a steady-state shock) will be given by the difference of their respective fluid energy equations, i.e. \citep{simakov17}
\begin{equation}     
\begin{split}
\frac{3}{2}\frac{\text{d}\left(T_2 - T_1\right)}{\text{d}x} = \underbrace{T_2\frac{\text{d}\left(\text{log}\ n_2\right)}{\text{d}x} - T_1\frac{\text{d}\left(\text{log}\ n_1\right)}{\text{d}x}}_\text{{\it PdV} terms} + \\ 
\underbrace{\frac{\pi_{2xx}}{n_2}\frac{\text{d}\left(\text{log}\ n_2\right)}{\text{d}x} - \frac{\pi_{1xx}}{n_1}\frac{\text{d}\left(\text{log}\ n_1\right)}{\text{d}x}}_\text{viscosity terms} + \\
\underbrace{\frac{1}{n_1u_1}\frac{\text{d}q_1}{\text{d}x} - \frac{1}{n_2u_2}\frac{\text{d}q_2}{\text{d}x}}_\text{heat flux terms} + \\
 \underbrace{\frac{Q_{2e}}{n_2u_2} - \frac{Q_{1e}}{n_1u_1}}_\text{electron-ion exch. terms} + \underbrace{\frac{Q_{21}}{n_2u_2} - \frac{Q_{12}}{n_1u_1}}_\text{ion-ion exch. terms},
\end{split}
\label{temp_diff_eq}
\end{equation}
where $\pi_{sxx}$, $Q_{ss'}$, and $q_s$ are the viscosity, energy exchange with species, $s'$, and heat flux of species, $s$, respectively. This equation shows that the ion temperature separation is driven by: 1) viscous heating; 2) ion-ion energy exchange; 3) electron-ion energy exchange; 4) ion heat flux; and 5) $PdV$ work. The ion-electron energy exchange terms are $\sim \sqrt{m_e/m_i}$ smaller than ion-ion terms, and thus may be ignored.
\newline
\indent
From Eq.\ (\ref{temp_diff_eq}), it is evident that the ion temperature separation is determined by a number of competing processes. In the next section, we will show from iFP simulations how the complex interplay of the terms in Eq.\ (\ref{temp_diff_eq}) is responsible for ion temperature separation.  

\subsection{Ion Temperature Separation from iFP}
\label{s:temp_sep_sim} 

\begin{figure}[thb]
\includegraphics[angle = 0, width = 1\columnwidth]{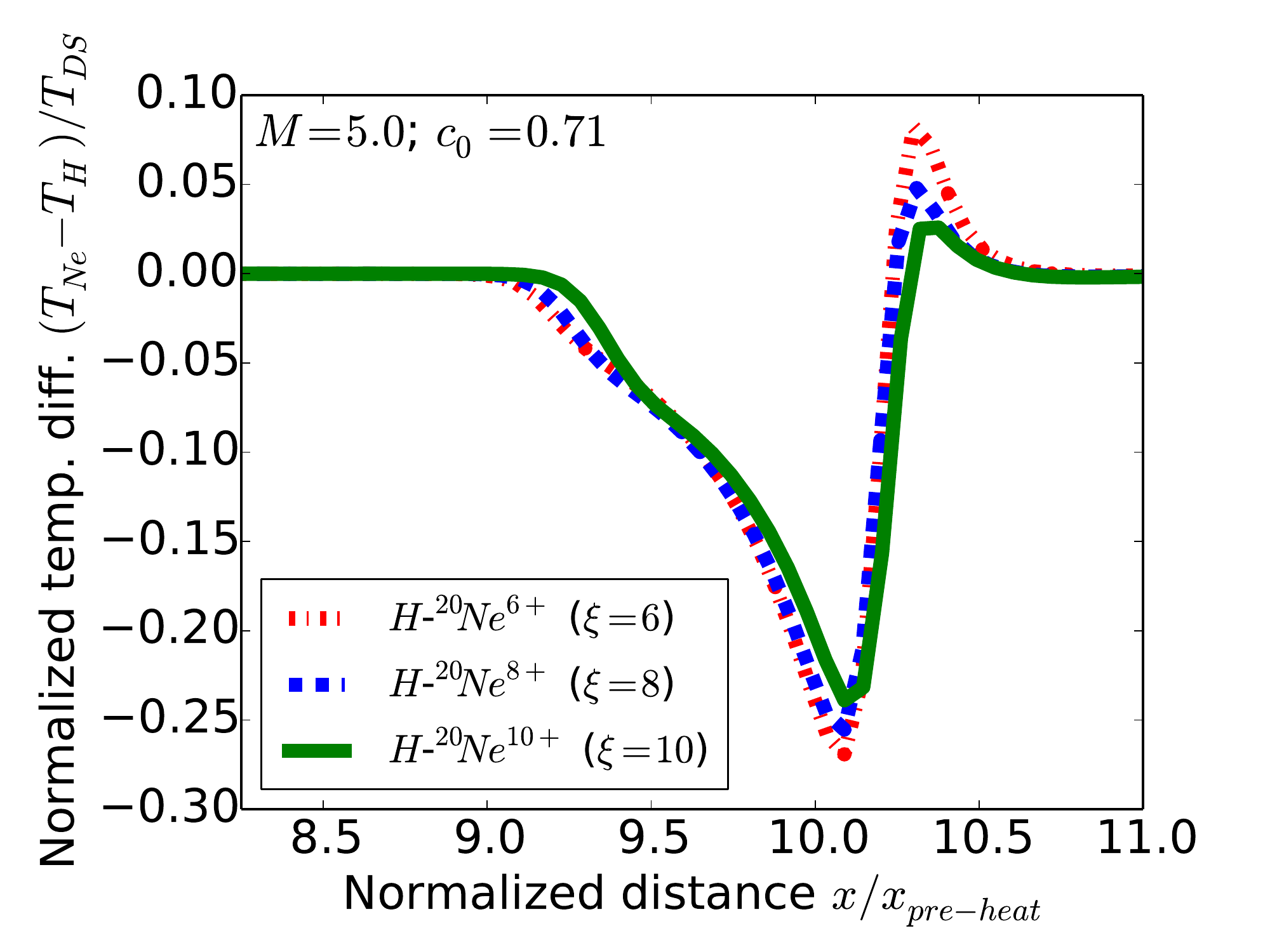}
\vskip-0.1cm
\caption{(Color online). Normalized ion temperature differences for H-Ne plasma mixtures with various charge ratios, $\xi$.}
\label{HNe_temp_diff_comp}
\end{figure}

\begin{figure}[thb]
\includegraphics[angle = 0, width = 1\columnwidth]{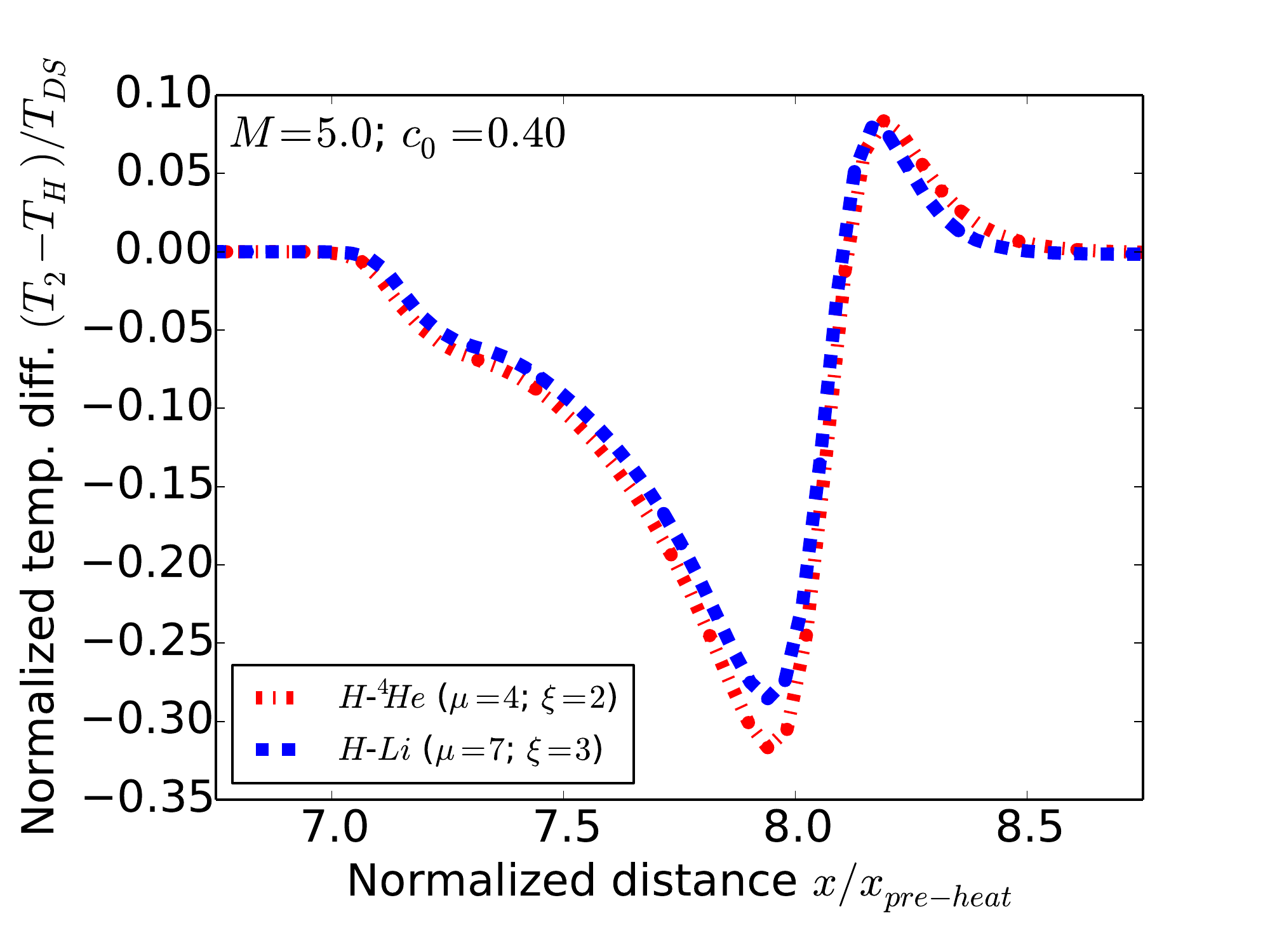}
\vskip-0.1cm
\caption{(Color online). Normalized ion temperature differences for various plasma mixtures. $T_2$ refers to the temperature of either $^4$He or Li. }
\label{temp_diff_mass}
\end{figure}

\begin{figure}[thb]
\includegraphics[angle = 0, width = 1\columnwidth]{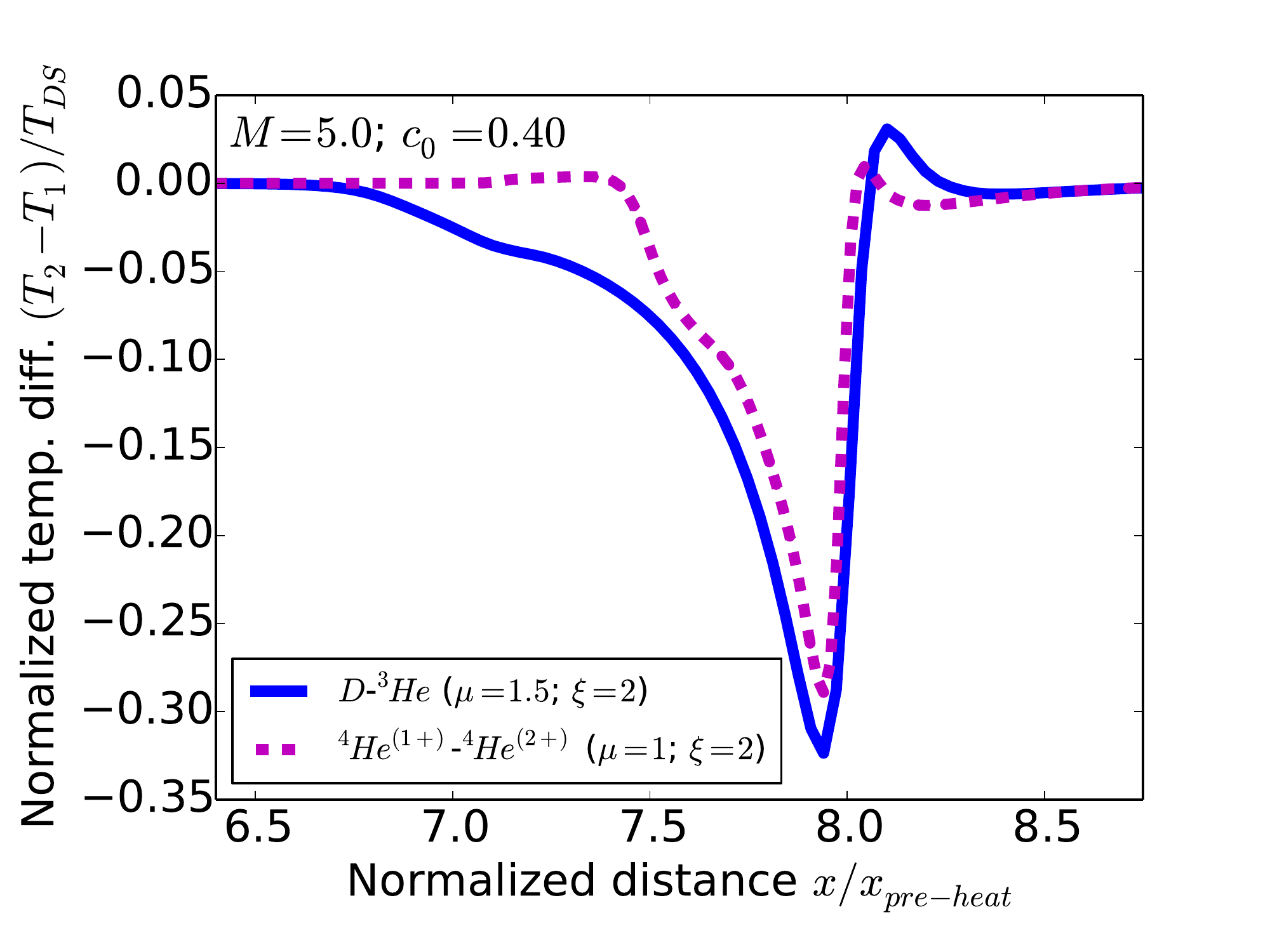}
\vskip-0.1cm
\caption{(Color online). Normalized ion temperature differences for two plasma mixtures. $T_2$ refers to the temperature of either $^3$He or $^4$He$^{(2+)}$.}
\label{temp_diff_mZ}
\end{figure}

Figures \ref{HNe_temp_diff_comp}, \ref{temp_diff_mass}, and \ref{temp_diff_mZ} show ion temperature separations obtained from iFP for various mixtures. The mixtures differ by $c_0$, $\mu$, and $\xi$,  but all of them display a higher temperature for the heavier species near the compression shock, and a higher temperature for the lighter species in the pre-heat layer.
\newline
\indent
We begin by exploring cases with large ion mass ratios, $\mu$. Figure \ref{HNe_temp_diff_comp} compares three different H-Ne mixtures ($M = 5$, $c_0 = 0.71$; $\mu = 20$). These mixtures differ by the degree of ionization of the neon ions. Notice that the pre-heat layer temperature differences are essentially the same. There is, however, considerable variation in the peak temperatures at the compression shock. Thus, these peak temperatures evidently strongly depend upon $\xi$. This is evidence that the peak temperature difference is driven principally by differences in ion viscosities and ion-ion energy exchange time-scales, which are both strong functions of $Z$. Note that the mixture with the largest charge ratio (H-$^{20}$Ne$^{10 +}$; $\mu = 20$; $\xi = 10$) exhibits the smallest peak (which continues throughout the thermal equilibration layer). With $Z = 10$, $^{20}$Ne$^{10 +}$ is strongly collisional. Hence, it is more effective at equalizing the inter-species temperatures. 
\newline
\indent
Figure \ref{temp_diff_mass} shows the ion temperature separations for H-$^4$He ($M = 5$; $c_0 = 0.40$; $\xi = 2$; $\mu = 4$) and H-Li ($M = 5$; $c_0 = 0.40$; $\xi = 3$; $\mu = 7$) mixtures. The two curves are very similar, and they show a lesser degree of variation near the peak temperature difference at the compression shock location than the H-Ne mixtures depicted in Fig.\ \ref{HNe_temp_diff_comp}. This is additional evidence that the peak temperature is very sensitive to $\xi$, since $\xi_{\text{H}-\text{Li}}/\xi_{\text{H}-^4\text{He}}$ is only slightly smaller than $\xi_{\text{H}-\text{Ne}^{10+}}/\xi_{\text{H}-\text{Ne}^{6+}}$. This, once again, suggests the ion-ion energy exchange time-scales are involved, since the self-collision frequency scales as $Z_i^4$, and inter-species collisions scale as $Z_i^2Z_j^2$ (i.e., the ion-ion energy exchange time-scales are very sensitive to $Z$).
\newline
\indent
Notice that the width of the curves, and their overall magnitudes, are very similar across all of the mixtures in Figs.\  \ref{HNe_temp_diff_comp} and \ref{temp_diff_mass}, despite the differences in $\mu \gg 1$, $\xi$, and $c_0$. This statement is particularly true in the pre-heat region. However, the widths do depend on $\mu$ when $\mu = \mathcal{O}(1)$. To see this, we consider mixtures with the same $\xi$, D-$^3$He ($\mu = 3/2$; $\xi = 2$) and $^4$He$^{(1+)}$-$^4$He$^{(2+)}$ ($\mu = 1$; $\xi = 2$) in Fig. \ \ref{temp_diff_mZ}. We see that the width of the ion temperature separation substantially differs between the mixtures in the pre-heat layer, indicating strong dependence upon $\mu$ (given nearly equal degrees of ion collisionality). Moreover, the fact that $\mu$ does not differ much between the two mixtures suggests that the D-$^3$He curve is broader because deuterons are significantly less massive than $^4$He ions.
\newline
\indent
This dependence upon $\mu$ can be accounted for in the heat flux terms in Eq.\ (\ref{temp_diff_eq}), since the ion heat flux is proportional to the ion mass. We will investigate this in the following sub-sections.  

\FloatBarrier

\subsection{Ion Heat Flux vs. Mach Number}
\label{s:heat_flux_M} 

\begin{figure}[thb]
\includegraphics[angle = 0, width = 1\columnwidth]{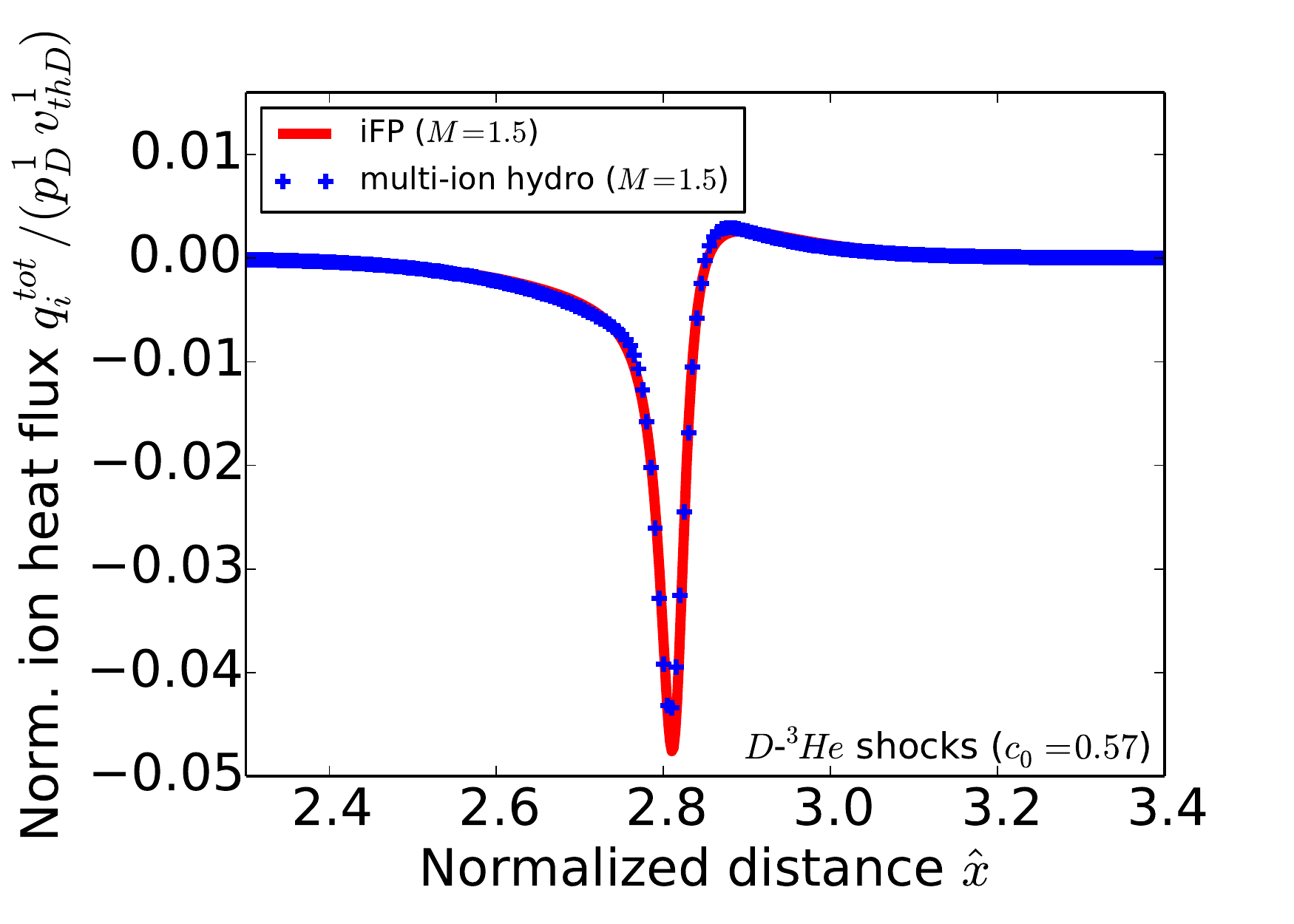}
\vskip-0.1cm
\caption{(Color online). Ion heat flux profiles from iFP and multi-ion hydro simulations for a weak D-$^3$He plasma shock ($M = 1.5$; $c_0 = 0.57$). Both heat fluxes are normalized to $p_\text{D}^1v_{th\text{D}}^1$, where $p_\text{D}^1$ and $v_{th\text{D}}^1$ are the downstream deuterium pressure and thermal velocity, respectively.}
\label{mach1.5_heat_flux}
\end{figure}

We will show in the next section that the ion temperature separation in the electron pre-heat layer is dominated by the ion heat flux terms in Eq.\ (\ref{temp_diff_eq}). However, first will we study the ion heat flux dependence on the Mach number, $M$. In particular, it will be useful to identify the manner in which the kinetic ion heat flux begins to deviate from the hydro equivalent as $M$ increases. We begin with the low Mach number limit. Figure \ref{mach1.5_heat_flux} shows the spatial profile of the ion heat flux from iFP (red solid line) for an $M = 1.5$ shock in D-$^3$He plasma. Also included in this figure is a profile (blue plus-sign line) from equivalent multi-ion hydro simulations.
\newline
\indent
Figure \ref{mach1.5_heat_flux} shows that the kinetic and the multi-ion hydro heat fluxes have very similar profiles, suggesting that the $M = 1.5$ shock is genuinely hydro-like. 
\begin{figure}[thb]
\includegraphics[angle = 0, width = 1\columnwidth]{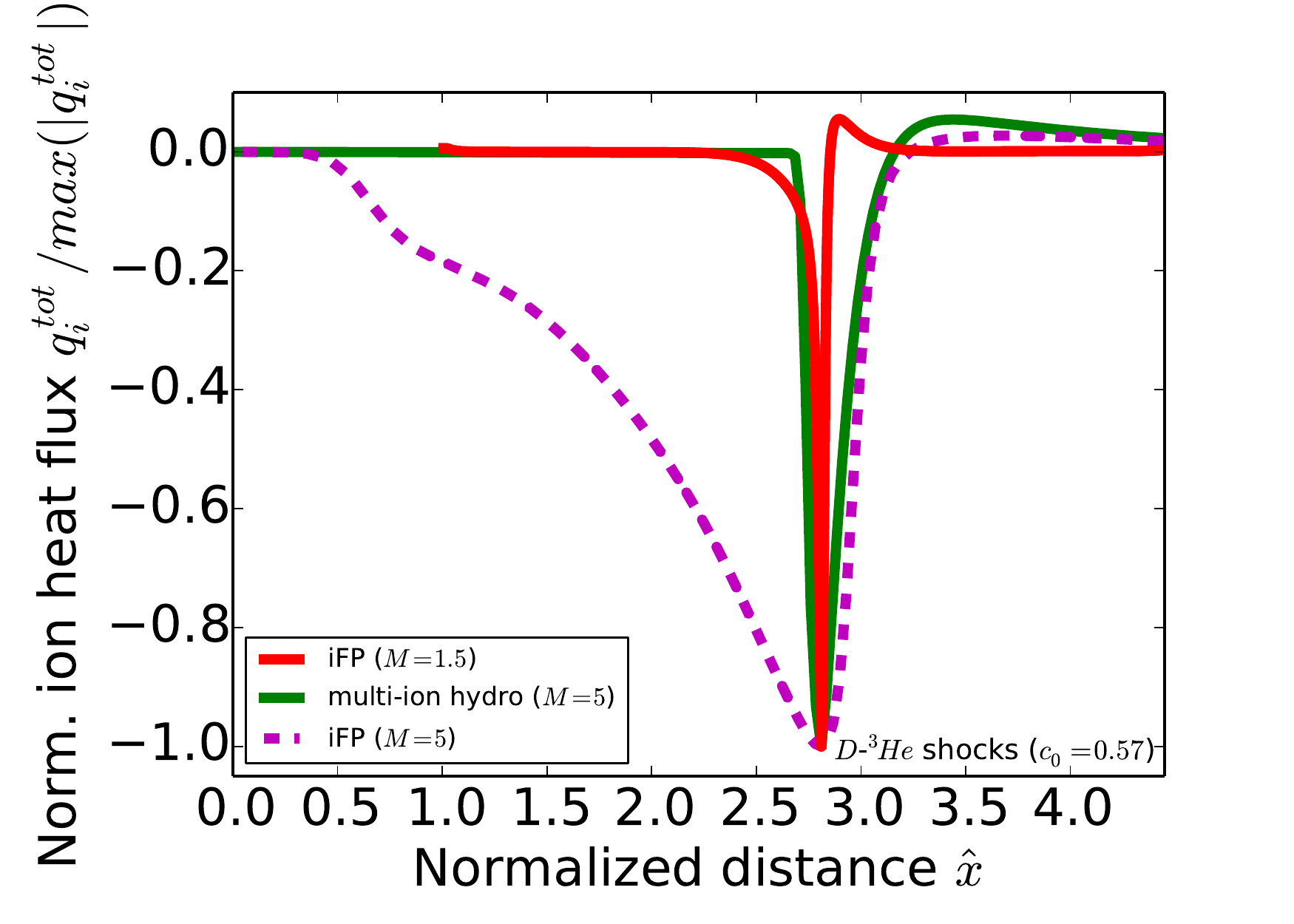}
\vskip-0.1cm
\caption{(Color online). Ion heat flux profiles (normalized to the maximum absolute value of the heat flux for each case) for D-$^3$He plasma shocks of varying strength ($c_0 = 0.57$).}
\label{mach5_heat_flux}
\end{figure}
\newline
\indent
The correspondence between hydro and kinetic predictions seen in Fig.\ \ref{mach1.5_heat_flux} should be contrasted with the larger departure displayed by strong shocks. As seen in Fig.\ \ref{mach5_heat_flux}, the multi-ion hydro heat flux for an $M = 5.0$ shock is small in the pre-heat layer of the shock, but the kinetic result shows considerable enhancement of the ion heat flux there. It follows that the ion temperature separation in the pre-heat layer is considerably suppressed in the hydro treatment. 
\begin{figure}[thb]
\includegraphics[angle = 0, width = 1\columnwidth]{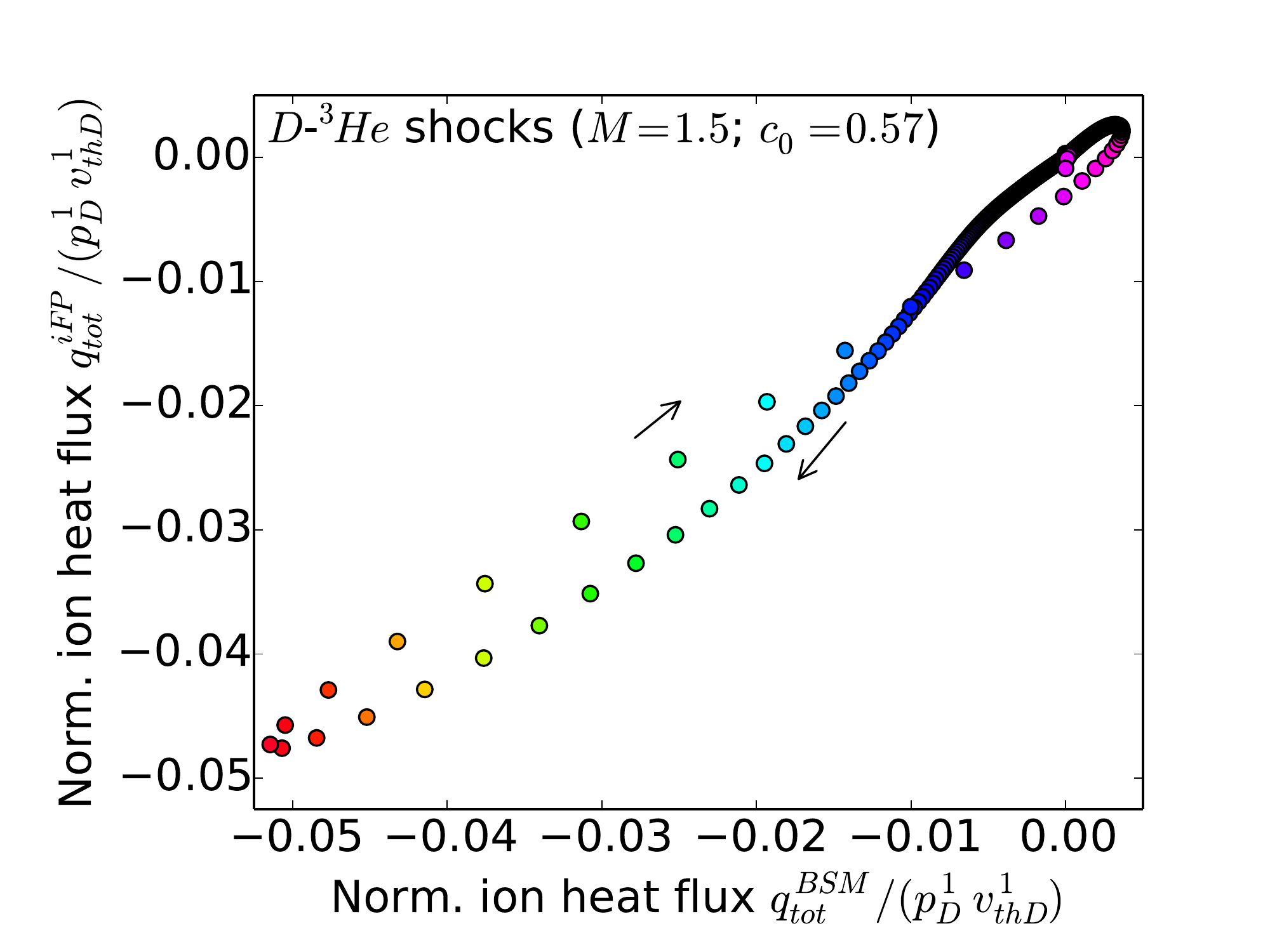}
\vskip-0.1cm
\caption{(Color online). The kinetic heat flux vs.\ the Braginskii-Simakov-Molvig \citep{simakov16, simakov16a} prediction for an $M = 1.5$ shock in D-$^3$He plasma.}
\label{mach15_iq}
\end{figure}
\newline
\indent
The degree to which intermediate-strength plasma shocks deviate from hydro predictions depends upon the plasma quantity in question, as certain quantities are more sensitive to the Knudsen number, $N_K$, than others. Since heat fluxes are predominately carried by particles with energies equal to several temperatures, they tend to be some of the most sensitive quantities. We show that an $M = 1.5$ shock's ion heat flux exhibits a subtle sensitivity to $N_K$ by plotting the multi-ion generalization of the Braginskii heat flux (i.e., the Braginskii-Simakov-Molvig, $BSM$, model \citep{simakov16, simakov16a} calculated from the kinetic temperature, densities, etc.) vs.\ the kinetic heat flux. As illustrated in Fig.\ \ref{mach15_iq}, the ion heat flux exhibits a hysteresis loop (instead of a straight line), \citep{negrete11} and is thus not a one-to-one function of hydrodynamic gradients (e.g.\ $\nabla T_i$, where $T_i$ is the ion temperature). Following colors from short-wavelength (violet) to long (blue), the points traverse the shock from the downstream to the upstream (i.e.\ the direction of the arrows). The narrow hysteresis loop is an indication of a moderately non-local nature of the heat flux.
\begin{figure}[thb]
\includegraphics[angle = 0, width = 1\columnwidth]{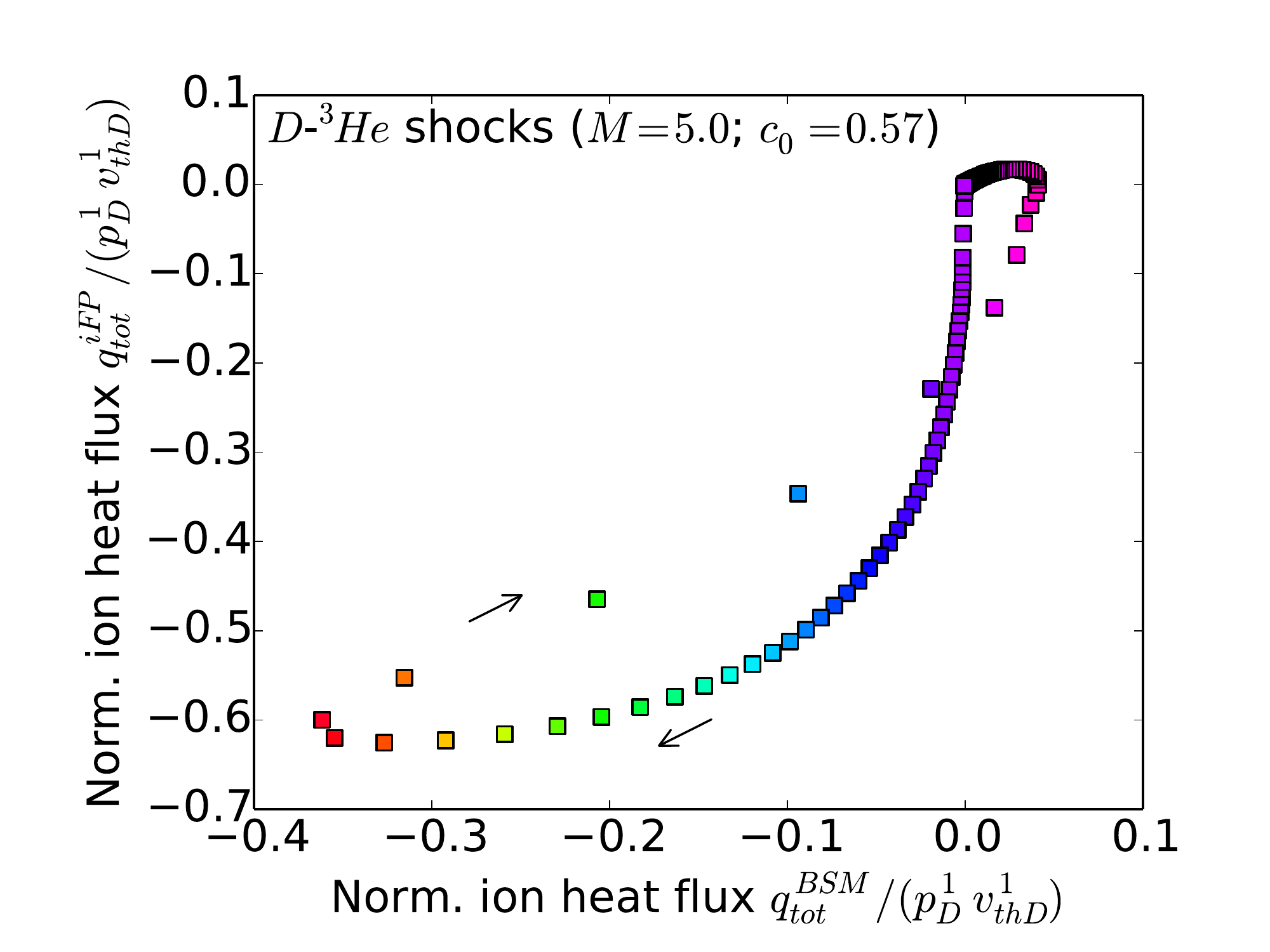}
\vskip-0.1cm
\caption{(Color online). The kinetic heat flux vs.\ the Braginskii-Simakov-Molvig \citep{simakov16, simakov16a} prediction for an $M = 5.0$ shock in D-$^3$He plasma.}
\label{mach5_iq}
\end{figure}
\newline
\indent
The hysteresis loop is more pronounced for higher Mach numbers. In Fig.\ \ref{mach5_iq}, we have plotted the same quantities for the $M = 5$ D-$^3$He shock from Fig. \ref{mach5_heat_flux}. The shape of this curve is notably more distorted, indicating strong non-local heat transport. Also note that the heat flux is an order of magnitude larger for $M = 5$ than for $M = 1.5$.

\FloatBarrier

\subsection{Ion Temperature Separation in the Electron Pre-Heat Layer}
\label{s:temp_sep_pre-heat} 

\begin{figure}[thb]
\includegraphics[angle = 0, width = 1\columnwidth]{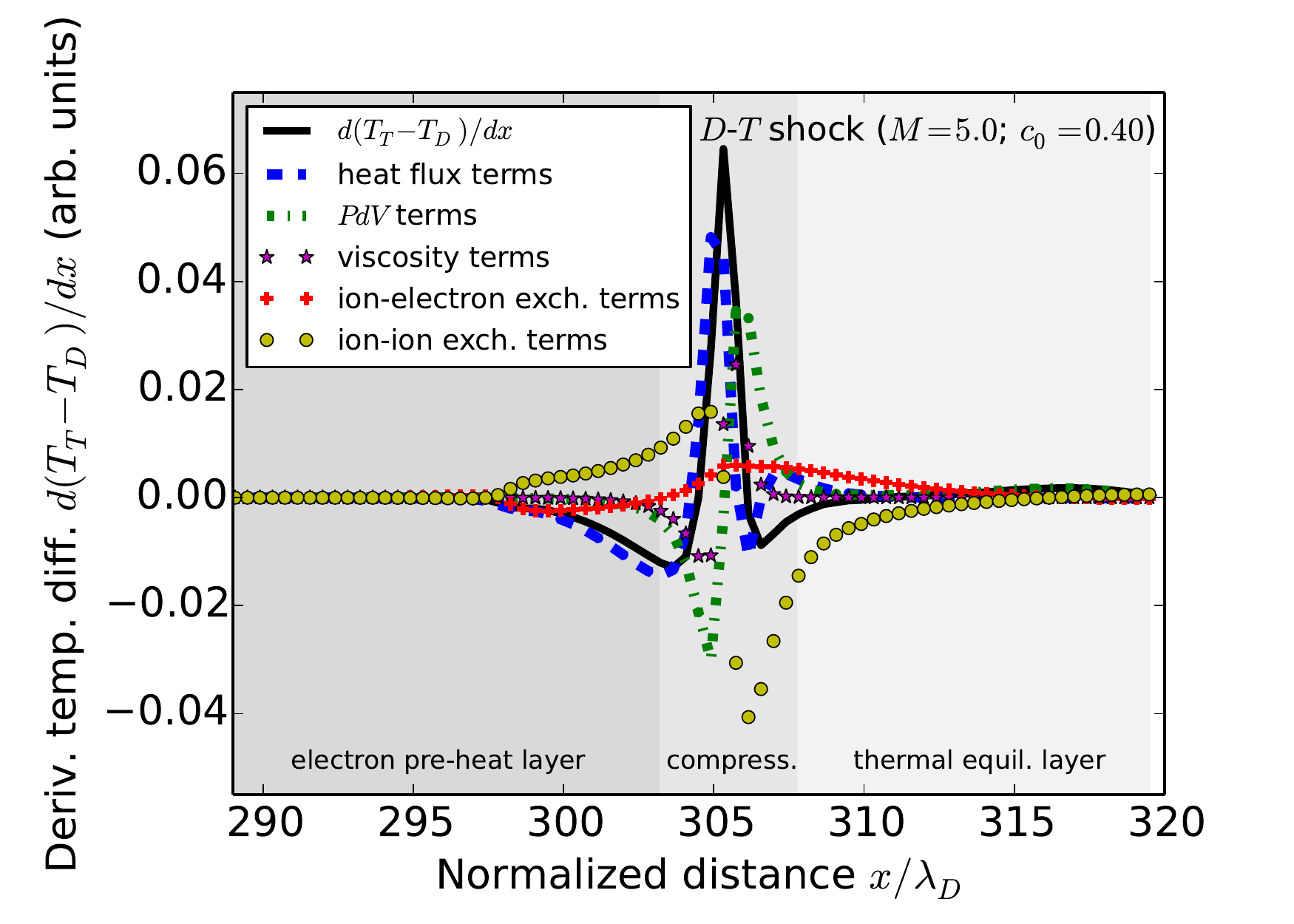}
\vskip-0.1cm
\caption{(Color online). The spatial derivative of the ion temperature separation for a D-T shock and its constituent parts from Eq.\ (\ref{temp_diff_eq}).}
\label{comp_deriv_temp_sep_ifp}
\end{figure}

\begin{figure}[thb]
\includegraphics[angle = 0, width = 1\columnwidth]{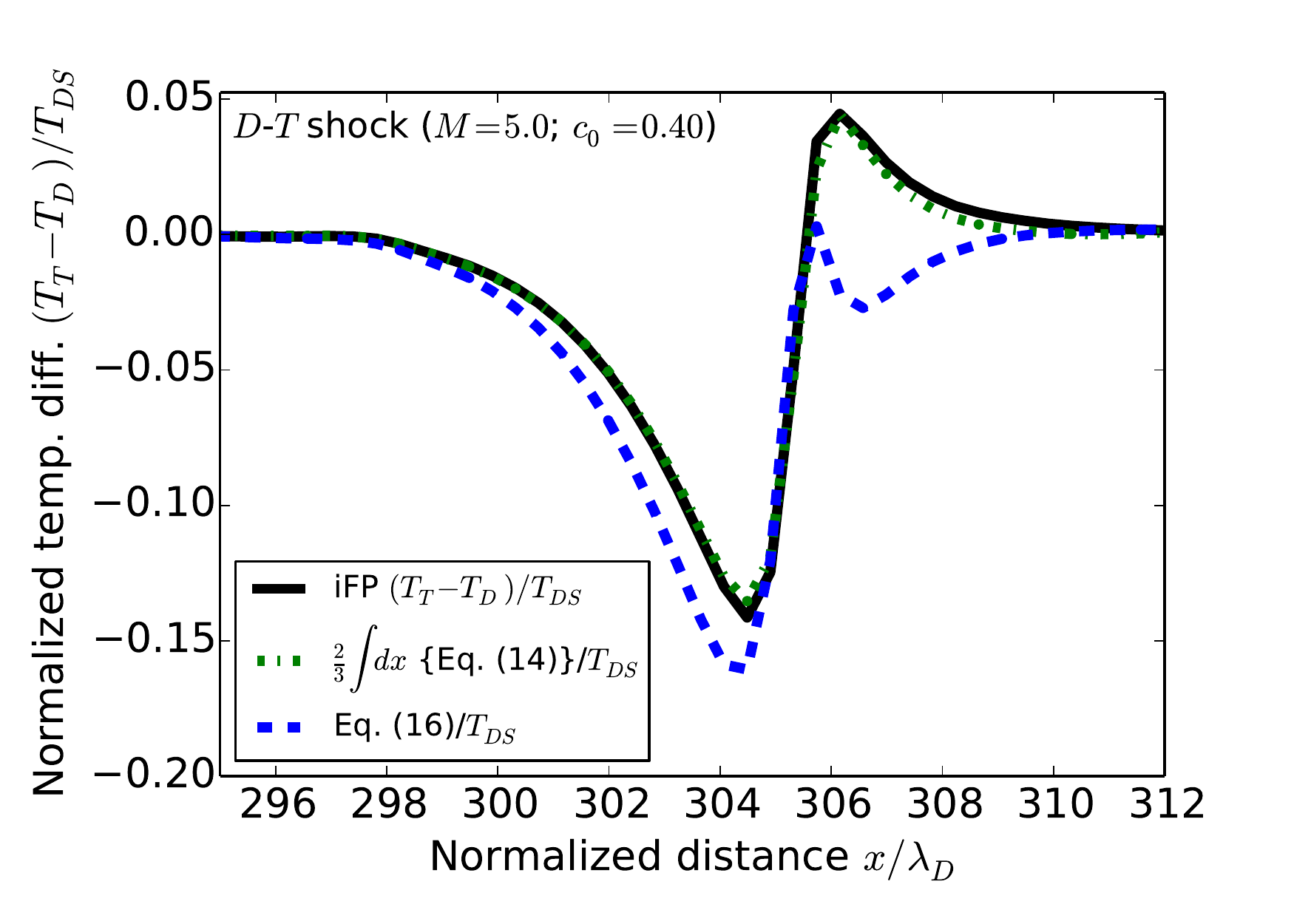}
\vskip-0.1cm
\caption{(Color online). Ion temperature separation for a D-T shock and the temperature separation predicted by Eqs.\ (\ref{temp_diff_eq}) and (\ref{temp_diff_sol}).}
\label{temp_sep_DT}
\end{figure}

In this sub-section, we show conclusively that the ion heat flux terms in Eq.\ (\ref{temp_diff_eq}) represent the most prominent contribution to the ion temperature separation in the electron pre-heat layer. To this end, it is helpful to analyze the separation's constituent parts. Figure \ref{comp_deriv_temp_sep_ifp} displays each of the terms from Eq.\ (\ref{temp_diff_eq}) for a D-T shock ($M = 5$; $\mu = 3/2$; $\xi = 1$). As before, we define the width of the ion compression shock as the distance over which the total mass density increases from 1.2 times its upstream value, $\rho_0$, to 0.9 times its downstream value, $\rho_1$. The term associated with the ion-ion energy exchange (yellow dots) is notably prominent throughout the shock, but it reaches its maximum (absolute) value in the ion compression layer. Most of the terms are, at least, slightly negative near the compression and thermal equilibration layers of the shock. This is why the heavier species has a higher temperature in the compression shock region. Additionally, the $PdV$ work term (green long-dash-short-dash line) and the viscosity term (magenta stars) are only large in the vicinity of the ion compression shock. Additionally, the ion-electron term (red plus-signs) is small throughout the shock (as expected). Finally, Fig.\ \ref{comp_deriv_temp_sep_ifp} clearly shows that the ion heat flux terms closely match the overall trend in $\text{d}(T_2 - T_1)/\text{d}x$ observed in the electron pre-heat layer.
\newline
\indent
Thus, we conclude that the ion heat flux terms from Eq.\ (\ref{temp_diff_eq}) dominate the ion temperature separation in the electron pre-heat layer. Hence, within this layer, Eq.\ (\ref{temp_diff_eq}) simplifies to:
\begin{equation}     
\frac{3}{2}\left[\frac{\text{d}\left(T_2 - T_1\right)}{\text{d}x}\right]_{pre-heat} \simeq \frac{1}{n_1^0u_0}\frac{\text{d}q_1}{\text{d}x} - \frac{1}{n_2^0u_0}\frac{\text{d}q_2}{\text{d}x},
\label{temp_diff_pre-heat}
\end{equation}
where we have used the mass continuity equation to find that $n_su_s = n_s^0u_0$. Eq.\ (\ref{temp_diff_pre-heat}) offers the exact solution:
\begin{equation}     
\left[T_2 - T_1\right]_{pre-heat} \simeq \frac{2}{3u_0}\left(\frac{q_1}{n_1^0} - \frac{q_2}{n_2^0}\right).
\label{temp_diff_sol}
\end{equation}
\newline
\indent
Figure \ref{temp_sep_DT} depicts a comparison of this solution with an iFP simulation result for a D-T plasma ($M = 5$, $c_0 = 0.40$).  Clearly, Eq.\ (\ref{temp_diff_sol}) (blue dashed line) captures the basic shape of the iFP curve, (black solid line) in the pre-heat layer. Additionally, the integral over all of the terms in Eq.\ (\ref{temp_diff_eq}) match very closely the iFP temperature difference. These terms (e.g., heat fluxes, viscosities, etc.) were calculated directly from iFP.  There is notable deviation between the heat-flux-only model and the total temperature difference around the compression shock and the thermal equilibration layer. This is expected, as the omitted terms from Eq.\ (\ref{temp_diff_eq}) in Eq.\ (\ref{temp_diff_pre-heat}) become important there.  

\FloatBarrier

\subsection{Ion Temperature Separation vs. Mach Number and Mass Fraction}
\label{s:temp_sep_mach_con} 

\begin{figure}[thb]
\includegraphics[angle = 0, width = 1\columnwidth]{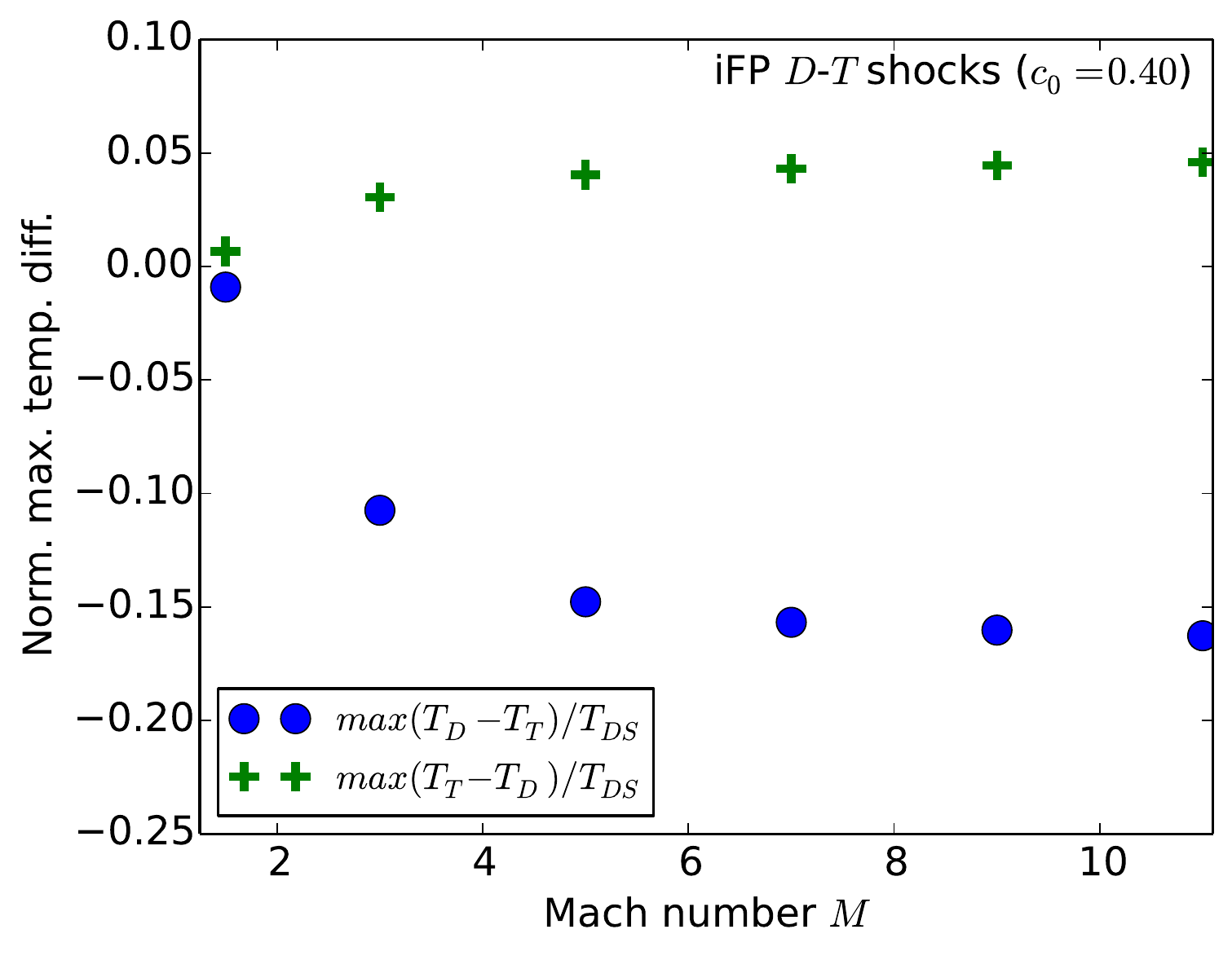}
\vskip-0.1cm
\caption{(Color online). The $max(T_T - T_D)$ and $max(T_D - T_T)$ vs.\ Mach number for D-T shocks with $c_0 = 0.40$.}
\label{temp_sep_mach_min}
\end{figure}

\begin{figure}[thb]
\includegraphics[angle = 0, width = 1\columnwidth]{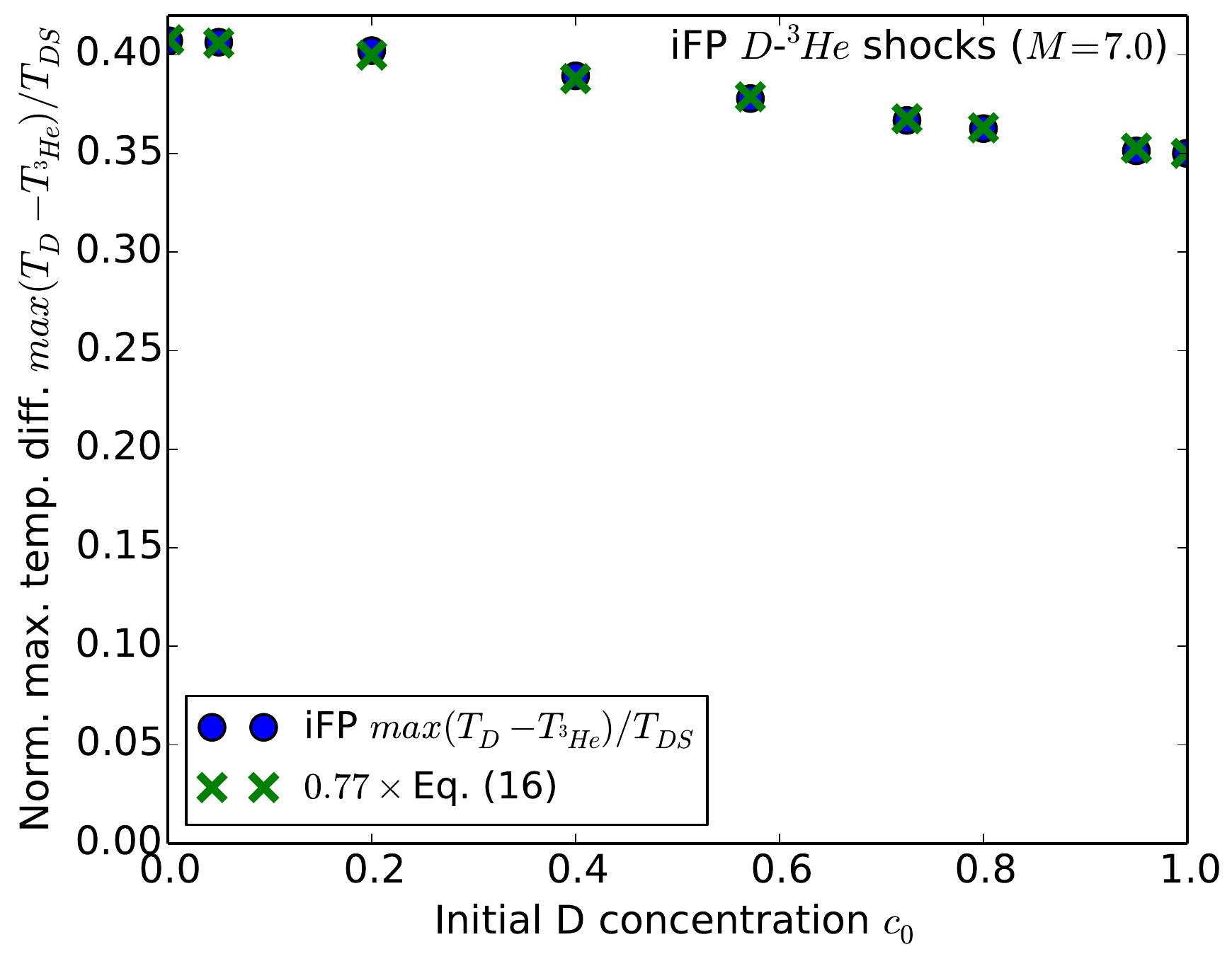}
\vskip-0.1cm
\caption{(Color online). The maximum difference in ion temperatures vs.\  $c_0$ for D-$^3$He shocks with $M = 7$.}
\label{temp_sep_c0_min}
\end{figure}

From the previous sub-sections, it is clear that ion temperature separation, $T_2 - T_1$, (where $m_2 > m_1$) displays two characteristic features: 1) a minimum value for which $T_1 > T_2$; and 2) a maximum value for which $T_2 > T_1$. In Fig.\ \ref{temp_sep_mach_min}, we have plotted these maximum and minimum values vs.\ Mach number for D-T shocks ($c_0 = 0.40$). Both the minimum and maximum ion temperature differences begin near zero for $M = 1.5$, and grow in their absolute values up to at least $M = 11$. The maximum and minimum differences appear to approach finite asymptotic values for $M \gg 1$. This is further evidence that kinetic effects saturate for $M \rightarrow \infty$. 
\newline
\indent
The effect of concentration is studied next. Figure \ref{temp_sep_c0_min} shows the maximum absolute D-$^3$He temperature separation, $max(T_{\text{D}} - T_{^\text{3He}})$, vs.\ $c_0$ for $M = 7$ (blue dots). The leftmost and rightmost iFP points are for $c_0 = 0.001$ and $c_0 = 0.999$, respectively. Evidently, the temperature separation is the greatest for a nearly pure $^3$He plasma (i.e., when the deuterium is very rarefied), but a finite amount of separation exists for all concentrations. This persistence of the temperature separation, independently of $c_0$, also occurs for weak shocks. \citep{simakov17} In all cases, the lighter species has a higher temperature than the heavier one, even when $n_D \sim 0$. 
\newline
\indent
Figure \ref{temp_sep_c0_min} also includes the $max(T_{\text{D}} - T_{^3\text{He}})$ given by Eq.\ (\ref{temp_diff_sol}). Up to a multiplicative factor ($\approx 0.77$), the predictions from Eq.\ (\ref{temp_diff_sol}) closely match the iFP results. The weak dependence of the $max(T_{\text{D}} - T_{^3\text{He}})$ on $c_0$ in Fig.\ \ref{temp_sep_c0_min} can be explained from Eq.\ (\ref{temp_diff_sol}) as follows. The heat flux for species, $s$ can be estimated as $q_s \sim p_{s} v_{th,s} \min(N_K^s,FL_s)$ with $p_s$, $v_{th,s} = \sqrt{2 T_s/m_s}$, $N_K^s$, and $FL_s$ the pressure, thermal speed, Knudsen number, and flux limiter, respectively. \citep{keenan17}  With $p_s \sim n_s^1T_{DS}$, Eq.\ (\ref{temp_diff_sol}) implies that:
\begin{equation}     
\begin{split}
max\left(T_\text{D} - T_{^3\text{He}}\right) \sim \frac{8\sqrt{2}}{3u_0}\left[\frac{\min(N_K^{^3\text{He}},FL_{^3\text{He}})}{\sqrt{m_{^3\text{He}}}} \right. \\
\left. - \frac{\min(N_K^{\text{D}},FL_{\text{D}})}{\sqrt{m_{\text{D}}}}\right],
\end{split}
\label{temp_diff_sol_next}
\end{equation}
where we have used $T_{DS} = 1$, and $n_s^1/n_s^0 \approx 4$. We expect that the Knudsen number (or flux limiter) pre-factors are only weakly dependent upon the ion concentration. Moreover, we are considering hydro-equivalent mixtures (in the upstream and downstream regions), and so $u_0$ is treated as a constant for a given Mach number. Given these considerations, Eq.\ (\ref{temp_diff_sol_next}) suggests that $max(T_{\text{D}} - T_{^3\text{He}})$ will only weakly depend upon $c_0$ through the Knudsen number, $N_K$ (or the flux limiter, $FL$).

\FloatBarrier

\section{Conclusions}
\label{s:disc}

We have studied collisional planar shocks in two-ion plasmas by considering their kinetic structure as a function of Mach number, lighter species concentration, ion mass ratio, and ion charge ratio. We first explored the impact of kinetic effects on the ion species relative concentration modifications by the shock. We showed that the lighter ion species concentration in a mixture is kinetically enhanced throughout the shock front, and particularly in the electron pre-heat layer. We showed that a semi-analytic theory, valid for $M \gg 1$, predicts the light species enrichment as a function of Mach number (given a multi-ion hydro treatment). This (multi-ion hydro) semi-analytic prediction scales as $M^4$ for $M \gg 1$, and we have found that the kinetic enrichment also obeys this $M^4$ scaling. When normalized to the ion shock width (which scales as $M^4$), the light species enrichment approaches a finite asymptotic limit as $M \rightarrow \infty$. The infinite Mach number kinetic deuterium enrichment (for a D-T plasma with $c_0 = 0.40$) is an order of magnitude greater than the multi-ion hydro one, indicating strong kinetic enhancement of the deuterium concentration in the shock front. Additionally, we showed that the integrated change (across the shock) in the light species concentration (i.e., the ion enrichment), when normalized to the ion shock width and the initial light species concentration, is a monotonically decreasing function of $c_0$. It peaks near $c_0 = 0$ (rarefied light species), and approaches zero as $c_0 \rightarrow 1$ (no heavy species). Finally, we found that the relative deuterium enrichment is well described by a characteristic enrichment distance, $x_\epsilon^{\text{D}} - Ax_\epsilon^{^3\text{He}}$ (where $x_\epsilon^{\text{D}}$ and $x_\epsilon^{^3\text{He}}$ are energy exchange distances for kinetic D and $^3$He ions, respectively, and $A$ is a constant).
\newline
\indent
Next, we explored ion temperature separation in strong shocks. It has been stated, given a hydrodynamic treatment, that the ion temperature within a shock should scale linearly with the ion mass. \citep{zeldovich67, rinderknecht15} We find that this is not the case anywhere within the shock structure. Rather, ion temperature separation in shocks is caused by a number of competing processes. In the pre-heat layer, the separation is dominated by differences in the kinetically enhanced ion heat fluxes; whereas ion viscosities, $PdV$ work, and thermal exchange play the most important roles in the compression and thermal equilibration layers of the shock. As a result of these processes, we find that the lighter species always has a higher temperature than the heavier species in the pre-heat layer. In contrast, we find that the heavier species always have a slightly greater temperature in the compression and thermal equilibration layers. 

\begin{acknowledgments}

{\it Acknowledgments}.--This work was supported by the Los Alamos National Laboratory LDRD Program, Metropolis Postdoctoral Fellowship for W.T.T., and used resources provided by the Los Alamos National Laboratory Institutional Computing Program. Work performed under the auspices of the U.S.\ Department of Energy National Nuclear Security Administration under Contract No.\ DE-AC52-06NA25396.

\end{acknowledgments}

\end{document}